\documentclass[onecolumn,unpublished,a4paper]{quantumarticle}
\usepackage[utf8]{inputenc}
\usepackage[backend=biber,style=alphabetic,maxbibnames=999]{biblatex}
\usepackage{amsmath}
\usepackage{amssymb}
\usepackage{amsthm}
\usepackage{amsfonts}
\usepackage[caption=false]{subfig}
\usepackage[colorlinks]{hyperref}
\usepackage[all]{hypcap}
\usepackage{tikz}
\usepackage{relsize}
\usepackage{color,soul}
\usepackage{capt-of}
\usepackage{mathtools}
\usepackage{float}
\usepackage[section]{placeins}
\usepackage{listings}
\usepackage[T1]{fontenc}  
\usetikzlibrary{decorations.pathreplacing}
\usepackage{braket}

\DeclareFixedFont{\ttb}{T1}{txtt}{bx}{n}{4}
\DeclareFixedFont{\ttm}{T1}{txtt}{m}{n}{4}
\definecolor{deepblue}{rgb}{0,0,0.5}
\definecolor{deepred}{rgb}{0.6,0,0}
\definecolor{deepgreen}{rgb}{0,0.5,0}
\newcommand\cppstyle{\lstset{
language=C++,
basicstyle=\ttm,
otherkeywords={uint8_t, __m256i, size_t, ASSERT_TRUE, EXPECT_TRUE, TEST, BENCHMARK},
keywordstyle=\ttb\color{deepblue},
emphstyle=\ttb\color{deepblue},
stringstyle=\color{deepgreen},
commentstyle=\fontfamily{txtt}\selectfont\color{gray},
showstringspaces=false,
literate={*}{{\char42}}1
         {-}{{\char45}}1
}}
\lstnewenvironment{cpp}[1][]
{\cppstyle\lstset{#1}}{}

\newcommand\pythonstyle{\lstset{
language=python,
basicstyle=\ttm,
morekeywords={assert,as,echo},
keywordstyle=\ttb\color{deepblue},
emphstyle=\ttb\color{deepblue},
stringstyle=\color{deepgreen},
commentstyle=\fontfamily{txtt}\selectfont\color{gray},
showstringspaces=false,
literate={*}{{\char42}}1
         {-}{{\char45}}1
}}
\lstnewenvironment{python}[1][]
{\pythonstyle\lstset{#1}}{}

\lstdefinestyle{stimcircuit}{
    language=python,
    basicstyle=\fontsize{6}{6}\selectfont\ttfamily,
    upquote=true,
    stepnumber=1,
    numbersep=8pt,
    showstringspaces=false,
    breaklines=true,
    frame=single,
    aboveskip=1.5em,
    belowskip=1.5em,
    commentstyle=\color{gray},
    classoffset=1,
    morekeywords={DETECTOR,OBSERVABLE_INCLUDE,rec},
    keywordstyle=\color{deepgreen},
    classoffset=2,
    morekeywords={H,R,MPP,M,RX,RY,MY,MX,SQRT\_X,XCY,XCZ,YCX},
    keywordstyle=\color{deepblue},
    classoffset=3,
    morekeywords={X_ERROR,DEPOLARIZE2,DEPOLARIZE1},
    keywordstyle=\color{red},
    classoffset=4,
    morekeywords={TICK,SHIFT_COORDS,QUBIT_COORDS},
    keywordstyle=\color{gray}
}

\renewcommand\thesection{\arabic{section}}

\theoremstyle{definition}

\theoremstyle{definition}

\theoremstyle{definition}

\newcommand{\eq}[1]{\hyperref[eq:#1]{Equation~\ref*{eq:#1}}}
\renewcommand{\sec}[1]{\hyperref[sec:#1]{Section~\ref*{sec:#1}}}
\DeclareRobustCommand{\app}[1]{\hyperref[app:#1]{Appendix~\ref*{app:#1}}}
\newcommand{\fig}[1]{\hyperref[fig:#1]{Figure~\ref*{fig:#1}}}
\newcommand{\tbl}[1]{\hyperref[tbl:#1]{Table~\ref*{tbl:#1}}}
\newcommand{\theoremref}[1]{\hyperref[theorem:#1]{Theorem~\ref*{theorem:#1}}}
\newcommand{\definitionref}[1]{\hyperref[definition:#1]{Definition~\ref*{definition:#1}}}

\begin{document}
\title{Inplace Access to the Surface Code Y Basis}

\date{\today}
\author{Craig Gidney}
\email{craig.gidney@gmail.com}
\affiliation{Google Quantum AI, Santa Barbara, California 93117, USA}

\begin{abstract}
In this paper, I cut the cost of Y basis measurement and initialization in the surface code by nearly an order of magnitude.
Fusing twist defects diagonally across the surface code patch reaches the Y basis in $\lfloor d/2 \rfloor + 2$ rounds, without leaving the bounding box of the patch and without reducing the code distance.
I use Monte Carlo sampling to benchmark the performance of the construction under circuit noise, and to analyze the distribution of logical errors.
Cheap inplace Y basis measurement reduces the cost of S gates and magic state factories, and unlocks Pauli measurement tomography of surface code qubits on space-limited hardware.
\end{abstract}

\emph{The source code that was written, the exact noisy circuits that were sampled, and the statistics that were collected as part of this paper are available at \href{https://doi.org/10.5281/zenodo.7487893}{doi.org/10.5281/zenodo.7487893}~\cite{gidneyybasisdata2022}.}

\maketitle

\section{Introduction}
\label{sec:introduction}

Because the surface code is a CSS code, its X and Z observables can be measured transversally.
The Y observable can't be measured transversally, and so must be reached by other means.
The cost of Y basis measurement is surprisingly relevant to the overall cost of surface code computations because, in many magic state factory designs~\cite{fowler2018latticesurgery,gidney2019catalyzeddistillation,litinski2018}, distilling a magic state involves hundreds of Y basis measurements.
If running Shor's algorithm costs a billion Toffoli gates~\cite{gidney2021factor,soeken2020improved} then, by implication, running Shor's algorithm also costs hundreds of billions of Y basis measurements.
The inefficiency of magic state factories magnifies the benefits of improving the cost of Y basis measurement.

\begin{figure}
    \centering
    \resizebox{\linewidth}{!}{
        \includegraphics{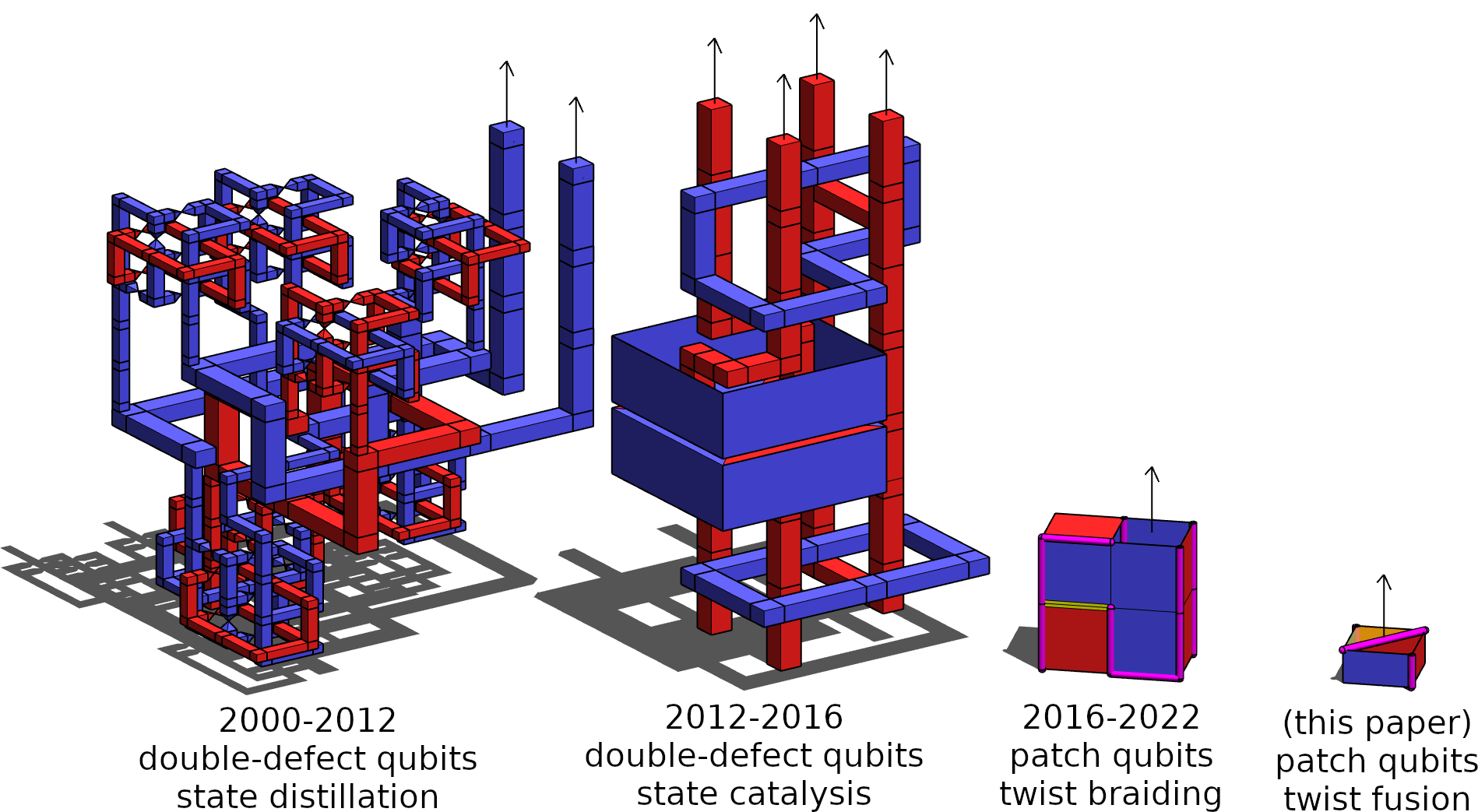}
    }
    \caption{
    To-scale defect diagrams of historical methods for planar initialization of $|i\rangle$ in the surface code, showing the huge improvement over time.
    See \tbl{defect_types} for a legend of defect types shown in the diagram.
    }
    \label{fig:historical_progression}
\end{figure}

Over the past decade, the cost of Y basis measurement and initialization in the surface code has dropped dramatically (see \fig{historical_progression}).
Originally, the only known way to reach the Y basis was by distillation of the $|i\rangle$ state~\cite{aliferis2005quantum,fowler2012surfacecodereview,fowler2012bridge}.
It was then realized that once an initial seed $|i\rangle$ state was prepared, it could catalyze the production of more $|i\rangle$ states~\cite{fowler2012surfacecodereview,gidney2017slightly}.
Later, it was shown that the $S$ gate could be performed by braiding twist defects, allowing $|i\rangle$ to be reached without distillation~\cite{brown2017surfacetwists}, with later realizations achieving a spacetime volume of $2d \times 2d \times d$~\cite{bombin2021logical,chamberland2022universal}.
This paper builds on these previous improvements and reduces the spacetime volume to $d \times d \times \frac{1}{2}d$, saving nearly another order of magnitude in the cost of reaching the Y basis.

A notable technique for reaching the Y basis, not mentioned in the previous paragraph, is folding the surface code patch~\cite{Kubica2015unfoldcolorcode,moussa2016folded}.
In a folded surface code patch, Y basis measurement can be performed in constant depth.
However, folding requires non-planar connectivity (or substantial routing overhead).
In this paper, I'm only interested in solutions with planar connectivity.
The reason for this focus is because I'm part of a superconducting qubit group, and superconducting qubit chips usually have planar connectivity~\cite{Andersen2020,googlerepcode2021,Zhao2022}.
That said, for completeness, I've included simulation results from folded surface codes in some figures.

The paper proceeds as follows.
In \sec{construction}, I'll explain the high level strategy and low level details behind my implementation of the Y basis measurement.
In \sec{benchmarking}, I'll describe and present statistics from various simulated experiments I performed, with the goal of understanding the quantity and causes of logical errors in the Y basis measurement.
In \sec{conclusion}, I'll summarize the results and discuss ideas for future work.
The paper also includes \app{defects}, which gives a summary of the surface code defects shown in diagrams, \app{noise_model}, which defines the superconducting-inspired noise model used in this paper, and \app{the_struggle}, which contains some remarks on how I managed the complexity of creating these circuits.

\section{Circuit Construction}
\label{sec:construction}

Surface code computation is topological in nature.
Constructions can be deformed without changing their function, because their function is defined by topological invariants such as whether two ``spacetime defects'' are connected.
(A spacetime defect is an exception to the normal background pattern of surface code.
See \app{defects} for definitions of the various spacetime defects.)
Thus, the first step to creating a surface code construction is to understand its topology.

\begin{figure}[h]
    \centering
    \resizebox{0.7\linewidth}{!}{
        \includegraphics{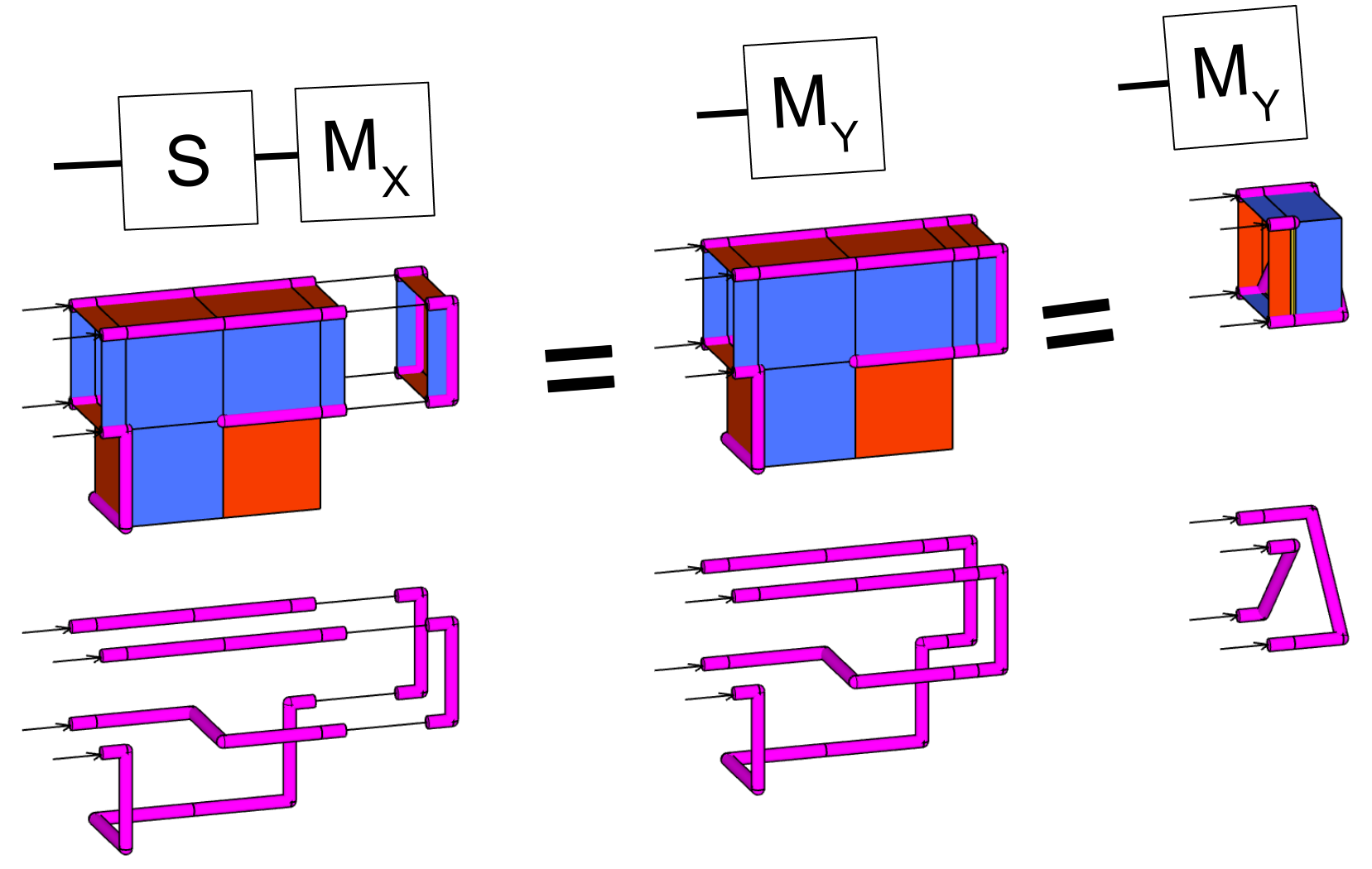}
    }
    \caption{
        Deriving that the Y basis measurement joins twist defects diagonally, by decomposing $M_Y$ into $S$ followed by $M_X$.
        See \tbl{defect_types} for a summary of the types of defects appearing in this diagram.
    }
    \label{fig:topological_decomposition}
\end{figure}

\begin{figure}
    \centering
    \resizebox{\linewidth}{!}{
    \includegraphics{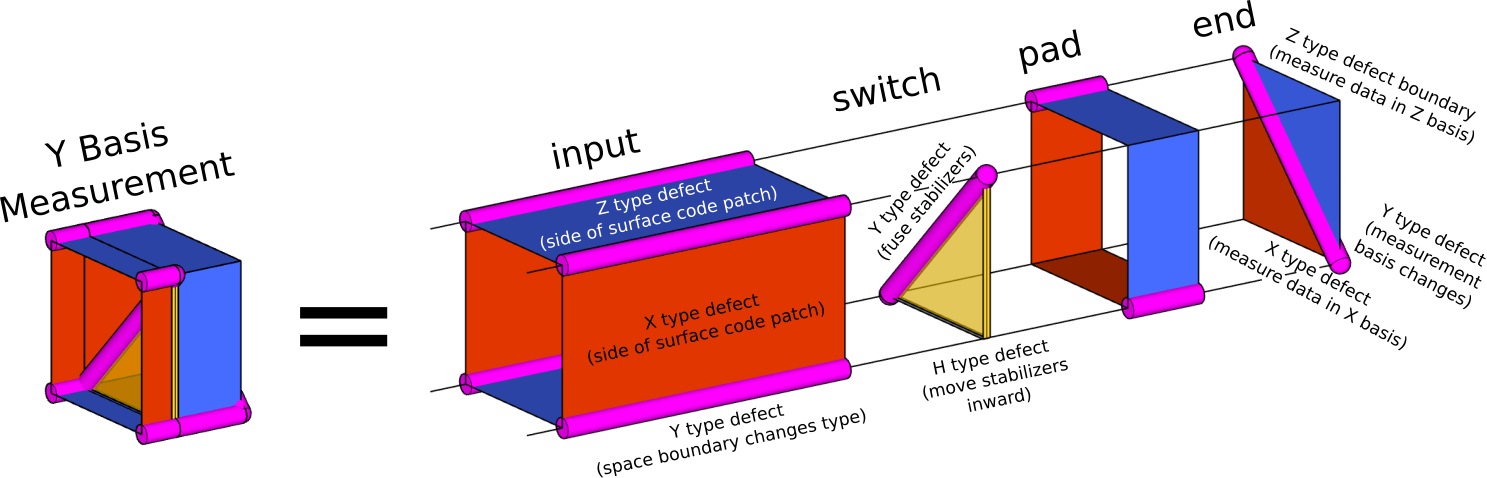}
    }
    \caption{
        Exploded defect diagram of an inplace Y basis measurement.
        See \tbl{defect_types} for a summary of the types of defects appearing in this diagram.
    }
    \label{fig:topological_explosion}
\end{figure}

One way to find a topological implementation of an operation is to decompose it into operations whose topology you already know, connect them together, and deform the combined object to simplify it.
Recall that a surface code patch has four twist defects, one at each corner~\cite{brown2017surfacetwists}.
The Y basis measurement can be decomposed into an S gate followed by an X basis measurement.
The S gate's topology is to exchange the two twists at the ends of an X type boundary of the surface code patch~\cite{brown2017surfacetwists}.
The X basis measurement's topology is to pair twists that share a Z type boundary, and join their ends together forming a single curve.
The topological result of composing these two operations also joins two twist defects, but it joins the twist defects across the diagonals of the patch.
Thus, the topology of the Y basis measurement is to join twists diagonally (see \fig{topological_decomposition} and \fig{topological_explosion}).

In previous work, measuring a surface code qubit in the Y basis required increasing the size of the patch, in order to move twists without letting them get too close to each other~\cite{brown2017surfacetwists,bombin2021logical,chamberland2022universal}.
My goal for this paper was to move the twists without resizing the patch.
This requires the twists to travel diagonally across the patch, without compromising the code distance of the circuit.

\begin{figure}
    \centering
    \resizebox{\linewidth}{!}{
    \includegraphics{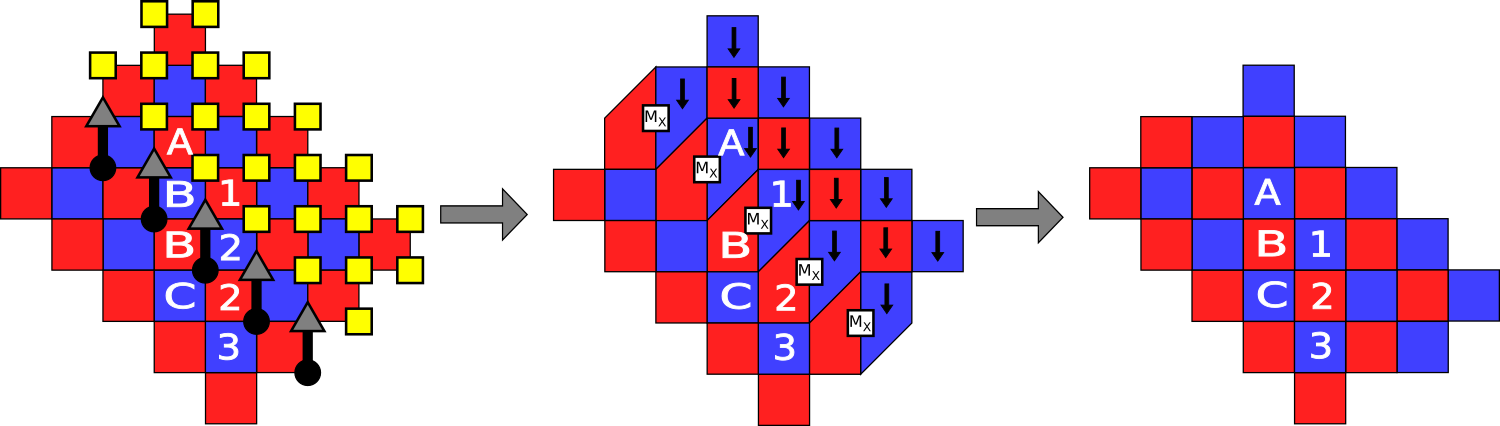}
    }
    \caption{
        An example of moving a twist diagonally through the bulk of the surface code.
        Boundaries not included.
        Simpler than the actual construction used by this paper (see \fig{patch_diagonal_twist} and \fig{transition_detector_slices}).
        The black circles connected to gray triangles are controlled-Y operations.
        The small yellow squares are Hadamard operations.
        The red shapes are X stabilizers.
        The blue shapes are Z stabilizers.
        \href{https://algassert.com/crumble\#circuit=Q(0,0)0;Q(0,1)1;Q(0,2)2;Q(0,3)3;Q(0,4)4;Q(0,5)5;Q(0,6)6;Q(0.5,0.5)7;Q(0.5,1.5)8;Q(0.5,2.5)9;Q(0.5,3.5)10;Q(0.5,4.5)11;Q(0.5,5.5)12;Q(1,0)13;Q(1,1)14;Q(1,2)15;Q(1,3)16;Q(1,4)17;Q(1,5)18;Q(1,6)19;Q(1.5,0.5)20;Q(1.5,1.5)21;Q(1.5,2.5)22;Q(1.5,3.5)23;Q(1.5,4.5)24;Q(1.5,5.5)25;Q(2,0)26;Q(2,1)27;Q(2,2)28;Q(2,3)29;Q(2,4)30;Q(2,5)31;Q(2,6)32;Q(2.5,0.5)33;Q(2.5,1.5)34;Q(2.5,2.5)35;Q(2.5,3.5)36;Q(2.5,4.5)37;Q(2.5,5.5)38;Q(3,0)39;Q(3,1)40;Q(3,2)41;Q(3,3)42;Q(3,4)43;Q(3,5)44;Q(3,6)45;Q(3.5,0.5)46;Q(3.5,1.5)47;Q(3.5,2.5)48;Q(3.5,3.5)49;POLYGON(0,0,1,0.5)3_16_17_4;POLYGON(0,0,1,0.5)15_28_29_16;POLYGON(0,0,1,0.5)5_18_19_6;POLYGON(0,0,1,0.5)17_30_31_18;POLYGON(0,0,1,0.5)29_42_43_30;POLYGON(0,0,1,0.5)31_44_45_32;POLYGON(0,0,1,0.5)1_14_15_2;POLYGON(0,0,1,0.5)13_26_27_14;POLYGON(0,0,1,0.5)27_40_41_28;POLYGON(1,0,0,0.5)16_29_30_17;POLYGON(1,0,0,0.5)2_15_16_3;POLYGON(1,0,0,0.5)18_31_32_19;POLYGON(1,0,0,0.5)4_17_18_5;POLYGON(1,0,0,0.5)28_41_42_29;POLYGON(1,0,0,0.5)30_43_44_31;POLYGON(1,0,0,0.5)14_27_28_15;POLYGON(1,0,0,0.5)0_13_14_1;POLYGON(1,0,0,0.5)26_39_40_27;TICK;R_8_22_10_12_24_34_36_38_20;RX_9_21_11_23_25_35_37_7_33;MARKX(0)23;MARKX(1)21;MARKZ(0)22;MARKZ(2)24;TICK;CX_1_8_21_14_9_2_15_22_3_10_23_16_11_4_17_24_5_12_25_18_37_30_31_38_27_34_35_28_29_36_7_0_13_20_33_26;TICK;CX_21_27_2_8_9_15_16_22_23_29_4_10_11_17_18_24_25_31_6_12_37_43_32_38_28_34_35_41_30_36_7_13_14_20_33_39;TICK;CX_21_15_14_8_9_3_28_22_23_17_16_10_25_19_18_12_11_5_30_24_44_38_37_31_40_34_42_36_35_29_7_1_26_20_33_27;TICK;CX_15_8_21_28_9_16_29_22_17_10_23_30_11_18_31_24_19_12_25_32_37_44_45_38_41_34_35_42_43_36_7_14_27_20_33_40;TICK;M_8_22_10_12_24_38_34_36_20;MX_9_21_11_23_25_37_35_7_33;MARKX(0)23;MARKX(1)21;MARKZ(0)22;MARKZ(2)24;TICK;CY_2_1_16_15_30_29_44_43;H_40_42_41_27_28_14_0_13_26_39;TICK;MX_15_1_29_43;SWAP_27_34_28_35_14_21_0_7_13_20_26_33_39_46_40_47_41_48_42_49;MARKX(0)15;MARKX(1)15;TICK;SWAP_34_28_35_29_21_15_7_1_20_14_33_27_46_40_47_41_48_42_49_43;TICK;R_8_22_10_12_24_34_36_38_20;RX_9_21_11_23_25_35_37_7_33;MARKX(0)23;MARKZ(1)22;MARKZ(2)24;TICK;CX_1_8_21_14_9_2_15_22_3_10_23_16_11_4_17_24_5_12_25_18_37_30_31_38_27_34_35_28_29_36_7_0_13_20_33_26;TICK;CX_21_27_2_8_9_15_16_22_23_29_4_10_11_17_18_24_25_31_6_12_37_43_32_38_28_34_35_41_30_36_7_13_14_20_33_39;TICK;CX_21_15_14_8_9_3_28_22_23_17_16_10_25_19_18_12_11_5_30_24_44_38_37_31_40_34_42_36_35_29_7_1_26_20_33_27;TICK;CX_15_8_21_28_9_16_29_22_17_10_23_30_11_18_31_24_19_12_25_32_37_44_45_38_41_34_35_42_43_36_7_14_27_20_33_40;TICK;M_8_22_10_12_24_38_34_36_20;MX_9_21_11_23_25_37_35_7_33;MARKX(0)23;MARKZ(1)22;MARKZ(2)24}{Click here to open a circuit with this transition in crumble.}
    }
    \label{fig:bulk_diagonal_twist}
\end{figure}

The first circuit I found for moving a twist diagonally is summarized in \fig{bulk_diagonal_twist}.
However, although that circuit is simple, it isn't efficient.
It forces the surface code cycle to wait, while the data qubits are moved using two layers of swap gates.
It would be far better to somehow merge the motion into the surface code cycle, to reduce the amount of extra noise incurred during the transition.

\begin{figure}
    \centering
    \resizebox{0.7\linewidth}{!}{
    \includegraphics{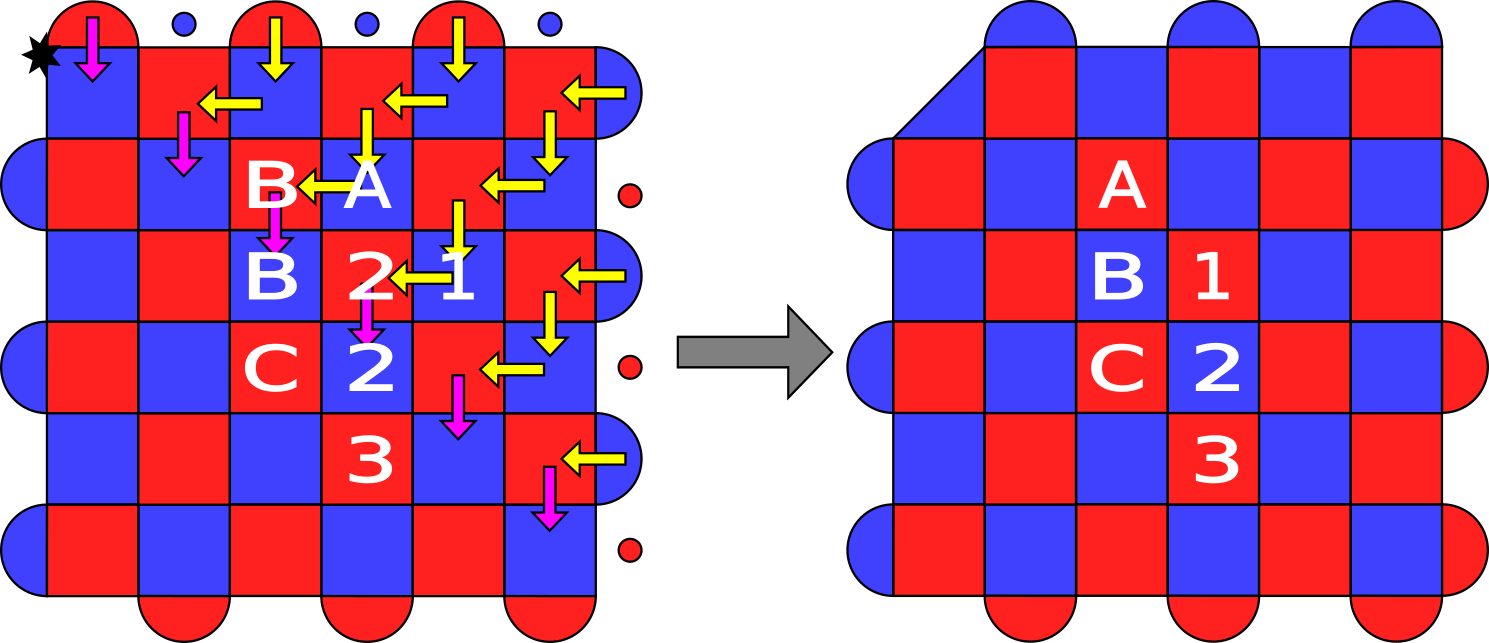}
    }
    \caption{
        Stabilizer flow during the key cycle of Y basis measurement.
        Red shapes are X stabilizers.
        Blue shapes are Z stabilizers.
        In the top right region X stabilizers take one step down while Z stabilizers take one step left.
        Transversal Hadamards turn the shifted X stabilizers into Z stabilizers, and vice versa.
        At the boundaries, new stabilizers are introduced to replace the ones that moved inward.
        At the center diagonal, the X stabilizers stepping down merge into the Z stabilizers sitting along the diagonal.
    }
    \label{fig:patch_diagonal_twist}
\end{figure}

\begin{figure}
    \centering
    \resizebox{\linewidth}{!}{
        \includegraphics{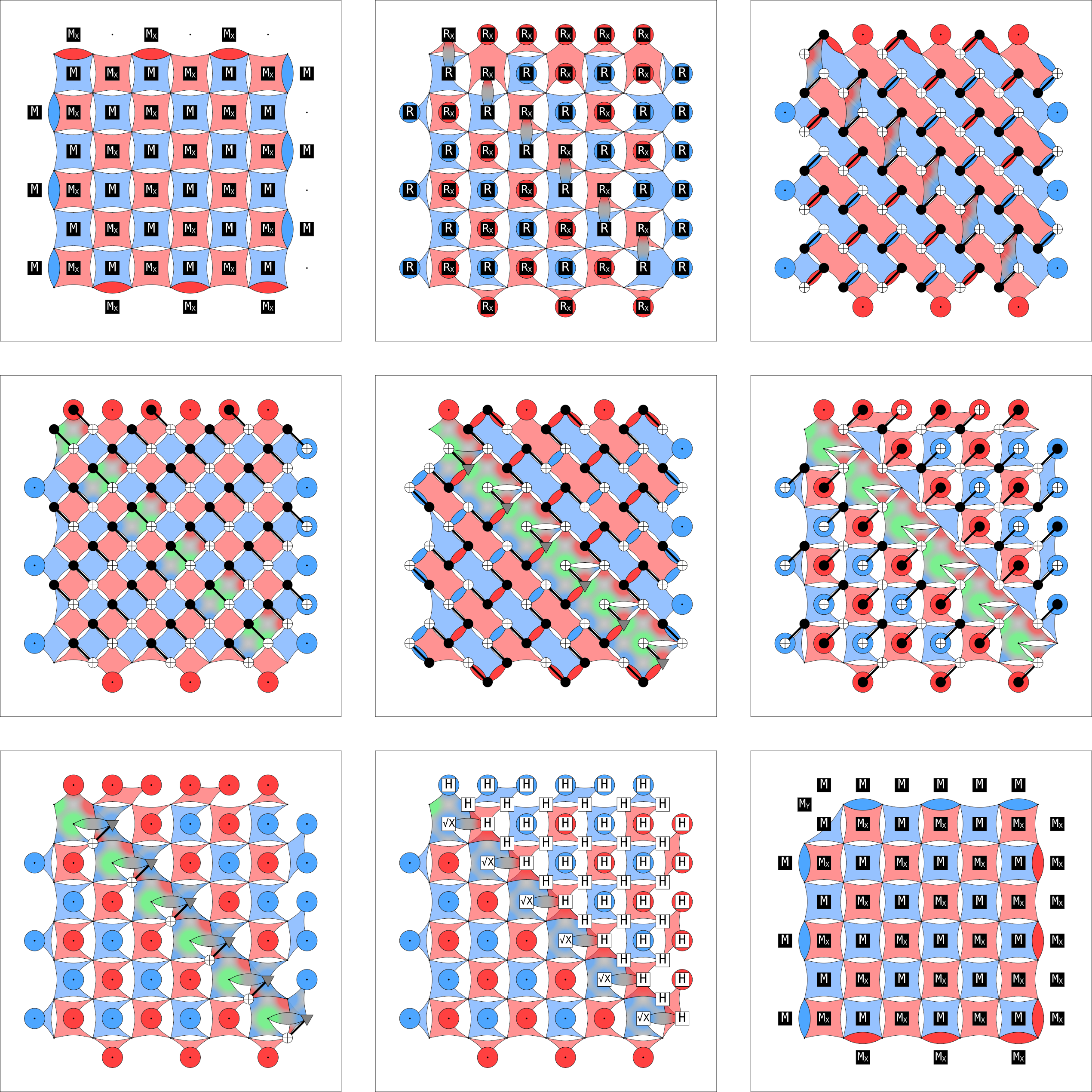}
    }
    \caption{
        A detector slice diagram of the surface code cycle where the logical qubit's Y observable is turned into a product of stabilizers, while continuing to measure the stabilizers of the patch.
        This cycle is the core driver of the logical Y basis measurement.
        Steps advance in reading order.
        The colored shapes shown behind each step are stabilizers of the state after the operations from that step have executed, using the color-to-pauli convention RGB=XYZ (red is X, green is Y, blue is Z).
        Crossed circles linked to gray triangles are X-controlled Y gates which apply a Y gate to the target when the control qubit is in the $|-\rangle$ state.
        This diagram shows the circuit before compilation into CZ gates.
        See also \fig{obs_slice}.
        \href{https://algassert.com/crumble\#circuit=Q(0,2)0;Q(0,4)1;Q(0,6)2;Q(0.5,0.5)3;Q(0.5,1.5)4;Q(0.5,2.5)5;Q(0.5,3.5)6;Q(0.5,4.5)7;Q(0.5,5.5)8;Q(0.5,6.5)9;Q(1,0)10;Q(1,1)11;Q(1,2)12;Q(1,3)13;Q(1,4)14;Q(1,5)15;Q(1,6)16;Q(1.5,0.5)17;Q(1.5,1.5)18;Q(1.5,2.5)19;Q(1.5,3.5)20;Q(1.5,4.5)21;Q(1.5,5.5)22;Q(1.5,6.5)23;Q(2,0)24;Q(2,1)25;Q(2,2)26;Q(2,3)27;Q(2,4)28;Q(2,5)29;Q(2,6)30;Q(2,7)31;Q(2.5,0.5)32;Q(2.5,1.5)33;Q(2.5,2.5)34;Q(2.5,3.5)35;Q(2.5,4.5)36;Q(2.5,5.5)37;Q(2.5,6.5)38;Q(3,0)39;Q(3,1)40;Q(3,2)41;Q(3,3)42;Q(3,4)43;Q(3,5)44;Q(3,6)45;Q(3.5,0.5)46;Q(3.5,1.5)47;Q(3.5,2.5)48;Q(3.5,3.5)49;Q(3.5,4.5)50;Q(3.5,5.5)51;Q(3.5,6.5)52;Q(4,0)53;Q(4,1)54;Q(4,2)55;Q(4,3)56;Q(4,4)57;Q(4,5)58;Q(4,6)59;Q(4,7)60;Q(4.5,0.5)61;Q(4.5,1.5)62;Q(4.5,2.5)63;Q(4.5,3.5)64;Q(4.5,4.5)65;Q(4.5,5.5)66;Q(4.5,6.5)67;Q(5,0)68;Q(5,1)69;Q(5,2)70;Q(5,3)71;Q(5,4)72;Q(5,5)73;Q(5,6)74;Q(5.5,0.5)75;Q(5.5,1.5)76;Q(5.5,2.5)77;Q(5.5,3.5)78;Q(5.5,4.5)79;Q(5.5,5.5)80;Q(5.5,6.5)81;Q(6,0)82;Q(6,1)83;Q(6,2)84;Q(6,3)85;Q(6,4)86;Q(6,5)87;Q(6,6)88;Q(6,7)89;Q(6.5,0.5)90;Q(6.5,1.5)91;Q(6.5,2.5)92;Q(6.5,3.5)93;Q(6.5,4.5)94;Q(6.5,5.5)95;Q(6.5,6.5)96;Q(7,1)97;Q(7,2)98;Q(7,3)99;Q(7,4)100;Q(7,5)101;Q(7,6)102;POLYGON(0,0,1,0.5)3_17_18_4;POLYGON(0,0,1,0.5)5_19_20_6;POLYGON(0,0,1,0.5)7_21_22_8;POLYGON(0,0,1,0.5)18_33_34_19;POLYGON(0,0,1,0.5)20_35_36_21;POLYGON(0,0,1,0.5)22_37_38_23;POLYGON(0,0,1,0.5)32_46_47_33;POLYGON(0,0,1,0.5)34_48_49_35;POLYGON(0,0,1,0.5)36_50_51_37;POLYGON(0,0,1,0.5)47_62_63_48;POLYGON(0,0,1,0.5)49_64_65_50;POLYGON(0,0,1,0.5)51_66_67_52;POLYGON(0,0,1,0.5)61_75_76_62;POLYGON(0,0,1,0.5)63_77_78_64;POLYGON(0,0,1,0.5)65_79_80_66;POLYGON(0,0,1,0.5)76_91_92_77;POLYGON(0,0,1,0.5)78_93_94_79;POLYGON(0,0,1,0.5)80_95_96_81;POLYGON(0,0,1,0.75)4_5;POLYGON(0,0,1,0.75)6_7;POLYGON(0,0,1,0.75)8_9;POLYGON(0,0,1,0.75)90_91;POLYGON(0,0,1,0.75)92_93;POLYGON(0,0,1,0.75)94_95;POLYGON(1,0,0,0.5)4_18_19_5;POLYGON(1,0,0,0.5)6_20_21_7;POLYGON(1,0,0,0.5)8_22_23_9;POLYGON(1,0,0,0.5)17_32_33_18;POLYGON(1,0,0,0.5)19_34_35_20;POLYGON(1,0,0,0.5)21_36_37_22;POLYGON(1,0,0,0.5)33_47_48_34;POLYGON(1,0,0,0.5)35_49_50_36;POLYGON(1,0,0,0.5)37_51_52_38;POLYGON(1,0,0,0.5)46_61_62_47;POLYGON(1,0,0,0.5)48_63_64_49;POLYGON(1,0,0,0.5)50_65_66_51;POLYGON(1,0,0,0.5)62_76_77_63;POLYGON(1,0,0,0.5)64_78_79_65;POLYGON(1,0,0,0.5)66_80_81_67;POLYGON(1,0,0,0.5)75_90_91_76;POLYGON(1,0,0,0.5)77_92_93_78;POLYGON(1,0,0,0.5)79_94_95_80;POLYGON(1,0,0,0.75)17_3;POLYGON(1,0,0,0.75)38_23;POLYGON(1,0,0,0.75)46_32;POLYGON(1,0,0,0.75)67_52;POLYGON(1,0,0,0.75)75_61;POLYGON(1,0,0,0.75)96_81;TICK;R_0_1_2_11_13_15_26_28_30_40_42_44_55_57_59_69_71_73_84_86_88_97_98_99_100_101_102;RX_10_12_14_16_24_25_27_29_31_39_41_43_45_53_54_56_58_60_68_70_72_74_82_83_85_87_89;MARKX(0)41;MARKX(1)70;MARKX(3)64_65_72_78_79;MARKZ(0)42;MARKZ(2)65_66_79_80_73;TICK;CX_10_3_12_5_14_7_16_9_25_18_27_20_29_22_39_32_41_34_43_36_45_38_54_47_56_49_58_51_68_61_70_63_72_65_74_67_83_76_85_78_87_80_4_11_6_13_8_15_19_26_21_28_23_30_33_40_35_42_37_44_48_55_50_57_52_59_62_69_64_71_66_73_77_84_79_86_81_88_91_97_93_99_95_101;TICK;CX_10_17_12_19_14_21_16_23_25_33_27_35_29_37_39_46_41_48_43_50_45_52_54_62_56_64_58_66_68_75_70_77_72_79_74_81_83_91_85_93_87_95_3_11_5_13_7_15_18_26_20_28_22_30_32_40_34_42_36_44_47_55_49_57_51_59_61_69_63_71_65_73_76_84_78_86_80_88_90_97_92_99_94_101;TICK;CX_17_25_33_41_46_54_48_56_62_70_64_72_75_83_77_85_79_87_91_98_93_100_95_102_24_32_40_47_53_61_55_63_69_76_71_78_82_90_84_92_86_94_12_4_14_6_16_8_27_19_29_21_31_23_43_35_45_37_58_50_60_52_74_66_89_81_5_0_7_1_9_2_20_13_22_15_36_28_38_30_51_44_67_59;XCY_11_18_26_34_42_49_57_65_73_80_88_96;TICK;CX_32_39_47_54_61_68_63_70_76_83_78_85_92_98_94_100_24_17_40_33_53_46_55_48_69_62_71_64_82_75_84_77_86_79_97_91_99_93_101_95_12_18_14_20_16_22_27_34_29_36_31_38_43_49_45_51_58_65_60_67_74_80_89_96_4_0_6_1_8_2_19_13_21_15_35_28_37_30_50_44_66_59;TICK;XCY_18_25_34_41_49_56_65_72_80_87_96_102;TICK;H_17_32_33_46_47_48_61_62_63_64_75_76_77_78_79_90_91_92_93_94_95_10_24_25_39_40_41_53_54_55_56_68_69_70_71_72_82_83_84_85_86_87_97_98_99_100_101_102;SQRT_X_11_26_42_57_73_88;TICK;M_0_1_2_10_11_13_15_24_26_28_30_39_40_42_44_53_55_57_59_68_69_71_73_82_84_86_88;MX_12_14_16_25_27_29_31_41_43_45_54_56_58_60_70_72_74_83_85_87_89_97_98_99_100_101_102;MY_3;MARKX(0)41_56;MARKX(2)87;MARKZ(0)42;MARKZ(1)71;MARKZ(2)73;MARKZ(3)71;TICK;POLYGON(0,0,1,0.5)17_18_4;POLYGON(0,0,1,0.5)5_19_20_6;POLYGON(0,0,1,0.5)7_21_22_8;POLYGON(0,0,1,0.5)18_33_34_19;POLYGON(0,0,1,0.5)20_35_36_21;POLYGON(0,0,1,0.5)22_37_38_23;POLYGON(0,0,1,0.5)32_46_47_33;POLYGON(0,0,1,0.5)34_48_49_35;POLYGON(0,0,1,0.5)36_50_51_37;POLYGON(0,0,1,0.5)47_62_63_48;POLYGON(0,0,1,0.5)49_64_65_50;POLYGON(0,0,1,0.5)51_66_67_52;POLYGON(0,0,1,0.5)61_75_76_62;POLYGON(0,0,1,0.5)63_77_78_64;POLYGON(0,0,1,0.5)65_79_80_66;POLYGON(0,0,1,0.5)76_91_92_77;POLYGON(0,0,1,0.5)78_93_94_79;POLYGON(0,0,1,0.5)80_95_96_81;POLYGON(0,0,1,0.75)4_5;POLYGON(0,0,1,0.75)6_7;POLYGON(0,0,1,0.75)8_9;POLYGON(0,0,1,0.75)32_17;POLYGON(0,0,1,0.75)61_46;POLYGON(0,0,1,0.75)90_75;POLYGON(1,0,0,0.5)4_18_19_5;POLYGON(1,0,0,0.5)6_20_21_7;POLYGON(1,0,0,0.5)8_22_23_9;POLYGON(1,0,0,0.5)17_32_33_18;POLYGON(1,0,0,0.5)19_34_35_20;POLYGON(1,0,0,0.5)21_36_37_22;POLYGON(1,0,0,0.5)33_47_48_34;POLYGON(1,0,0,0.5)35_49_50_36;POLYGON(1,0,0,0.5)37_51_52_38;POLYGON(1,0,0,0.5)46_61_62_47;POLYGON(1,0,0,0.5)48_63_64_49;POLYGON(1,0,0,0.5)50_65_66_51;POLYGON(1,0,0,0.5)62_76_77_63;POLYGON(1,0,0,0.5)64_78_79_65;POLYGON(1,0,0,0.5)66_80_81_67;POLYGON(1,0,0,0.5)75_90_91_76;POLYGON(1,0,0,0.5)77_92_93_78;POLYGON(1,0,0,0.5)79_94_95_80;POLYGON(1,0,0,0.75)38_23;POLYGON(1,0,0,0.75)67_52;POLYGON(1,0,0,0.75)96_81;POLYGON(1,0,0,0.75)91_92;POLYGON(1,0,0,0.75)93_94;POLYGON(1,0,0,0.75)95_96}{Click here to open this circuit in crumble.}
    }
    \label{fig:transition_detector_slices}
\end{figure}

The best circuit that I found for moving a twist diagonally is summarized in \fig{patch_diagonal_twist}, and shown exactly in \fig{transition_detector_slices}.
The intuition behind this circuit is based on understanding how topological defects can be arranged and implemented (see \app{defects}).
I knew twist defects must appear wherever domain wall defects end in the bulk, and I knew a domain wall defect could be implemented by using a walking surface code~\cite{mcewenmidoutsurfaces2023}.
Therefore I tried to produce a twist defect along the diagonal by running a normal surface code cycle in the bottom left of the patch while running a walking surface code cycle in the top right.

As noted in \tbl{defect_types}, what makes a domain wall a domain wall is that it maps X into Z and Z into X.
Hadamard gates exchange the X and Z basis, and so you might expect that transversal Hadamard gates are sufficient to implement a domain wall defect.
However, although the Hadamard-transformed stabilizers have undergone the right basis transition for a domain wall, they end up in the wrong location.
Just applying Hadamards produces stabilizers that don't match the surrounding checkerboard pattern of X and Z stabilizers.
Correcting this requires moving the stabilizers over by a little bit.
This can be done by using a walking surface code, and this is why I say the walking surface code cycle implements a domain wall defect.
The exact walking surface code cycle that I used ended up differing from the one presented in \cite{mcewenmidoutsurfaces2023}.
My cycle moves the surface code faster, but doesn't exchange the roles of the data qubits and the measurement qubits.

There were three main difficulties in getting the circuit details exactly right: finding operations that meshed the normal and walking surface codes at the seam along the diagonal, finding operations that held the boundaries in place as the walking surface code pulled stabilizers inward, and not losing code distance anywhere while doing these things.
I had no clever methodology for finding a satisfying circuit; I wrote a tool to make it easy to experiment and then experimented until I had a solution (see \app{the_struggle}).
In fact, when I initially started writing this paper, my best construction required two surface code cycles (instead of one) and reduced the code distance from $d$ to $d-1$.
I later found the circuit I'm presenting, which manages to merge all necessary qubit movement into a single working surface code cycle, while using only one additional layer of entangling gates, and while preserving the full code distance.
(I hoped there was even a way to avoid the additional layer of entangling gates, but I haven't been able to find one.)

Locally, the reason the circuit in \fig{transition_detector_slices} works is because it is measuring all the stabilizers of the input surface code configuration and preparing all the stabilizers of the output surface code configuration.
Globally, the reason this circuit works is because it changes the boundary types around the surface code patch from XZXZ to XXZZ.
These local and global changes have the combined effect of fault tolerantly mapping the Y observable to a known product of stabilizers.
A patch with XXZZ boundaries isn't capable of storing a logical qubit.
Instead, this patch has the property that, by multiplying all of the X basis stabilizers together, you can derive the value of a degenerate X observable running along the Z boundary.
This product of stabilizers is what the Y observable is mapped to.
By repeatedly measuring the stabilizers of the patch, you learn this product to arbitrarily high certainty~\cite{gidney2022stability}.
As will be shown in the next section, it's sufficient to measure these stabilizers $d/2$ times.
The measurement process is finished by destroying the patch by measuring all of its data qubits.
To maximize code distance, each data qubit is measured in the basis of its closest boundary.

I verified the correctness of my construction in a variety of ways.
First, locally, the key transition round is supposed to turn the Y observable into a product of measured stabilizers.
The code I wrote declares that this transformation is supposed to occur during this part of the circuit, and locally verifies it using the same method that's used to verify other stabilizer flows.
Second, globally, a Y basis measurement should have the property that it correctly measures the value produced by a Y basis initialization.
It's tempting to simply time reverse the Y basis measurement to produce a corresponding initialization, but this wouldn't detect mistakes such as accidentally measuring X instead of Y, so when verifying I used simple noiseless direct preparation of the observable and all stabilizers.
I also verified against a magic state injection circuit set to inject the $|i\rangle$ state.
Third, if the construction is working, it's supposed to have a code distance of $d$.
I checked the code distance using \href{https://github.com/quantumlib/Stim/blob/main/doc/python_api_reference_vDev.md#stim.Circuit.shortest_graphlike_error}{stim's graphlike distance search} and \href{https://github.com/quantumlib/Stim/blob/main/doc/python_api_reference_vDev.md#stim.Circuit.search_for_undetectable_logical_errors}{more general heuristic search}.
The shortest logical error found, within a $d \times d$ patch under circuit noise, uses $d$ physical error mechanisms.
The construction achieves the full code distance of the patch.

\begin{figure}
    \centering
    \resizebox{\linewidth}{!}{
        \includegraphics{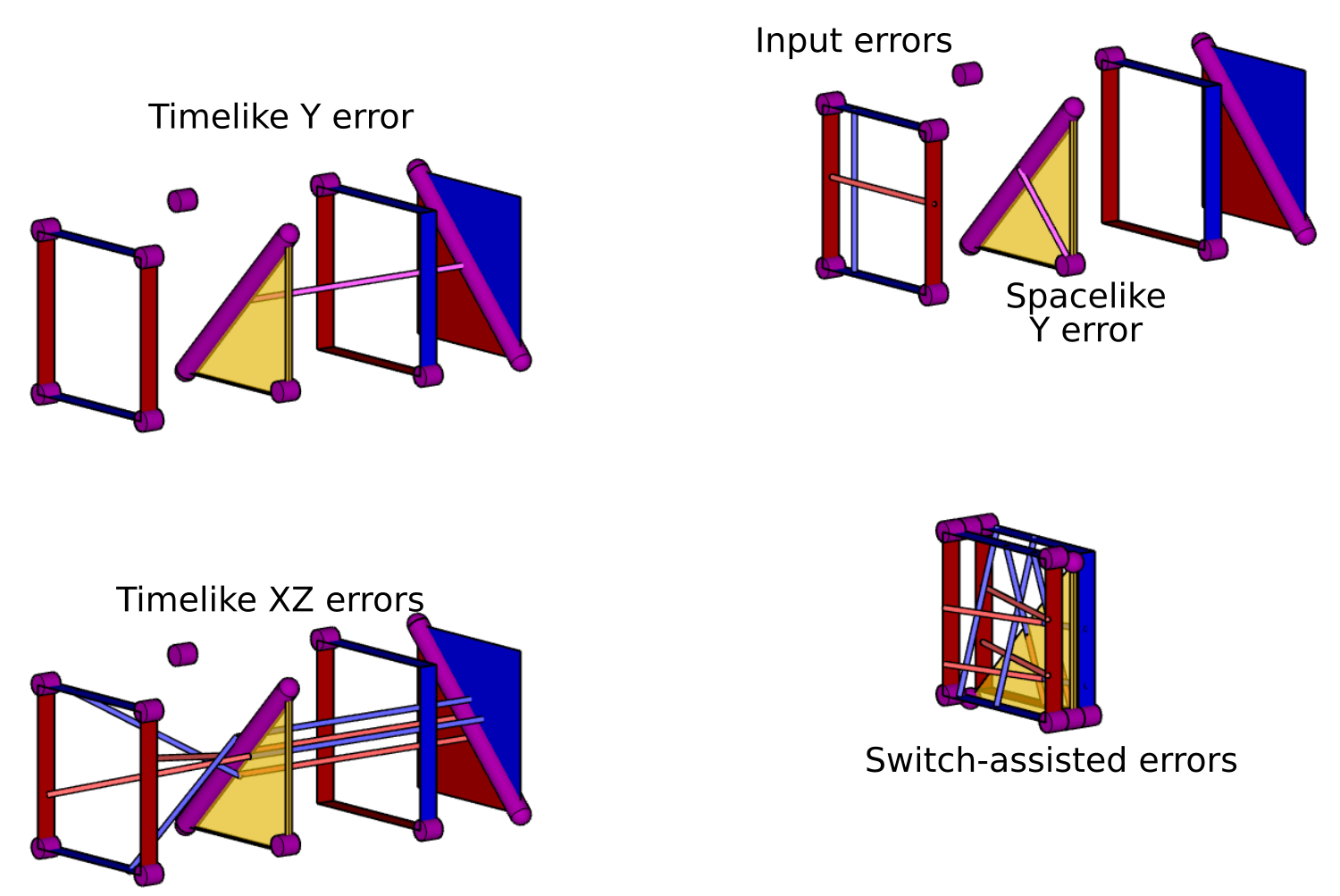}
    }
    \caption{
        Logical error mechanisms that prevent the construction from getting smaller, shown as small bright cylinders traveling between darkened defects.
        ``Input errors'' are errors that occur while the patch is a normal logical qubit patch, before the switch to a degenerate patch.
        ``Spacelike Y errors'' are error chains traveling from a corner twist to the crossing twist during the transition round.
        ``Timelike Y errors'' are error chains traveling from the twist defect in the transition round to defects in the final round.
        ``Timelike XZ errors'' are error chains traveling from the X and Z boundaries in the transition round to defects in the final round, that become topologically non-trivial due to traveling around the twist defect.
        ``Switch-assisted errors'' are error chains traveling from one boundary to another during the transition round.
        The timelike errors are what determine how fast the logical Y basis measurement can be performed, and also the best basis for each data qubit measurement.
        If a timelike error is a problem, it can be trivially mitigated by adding more padding rounds.
        Error chains travelling during the transition round are more difficult to mitigate without resizing the patch, requiring careful ordering of operations.
    }
    \label{fig:logical_error_paths}
\end{figure}

\begin{figure}
    \centering
    \resizebox{\linewidth}{!}{
        \includegraphics{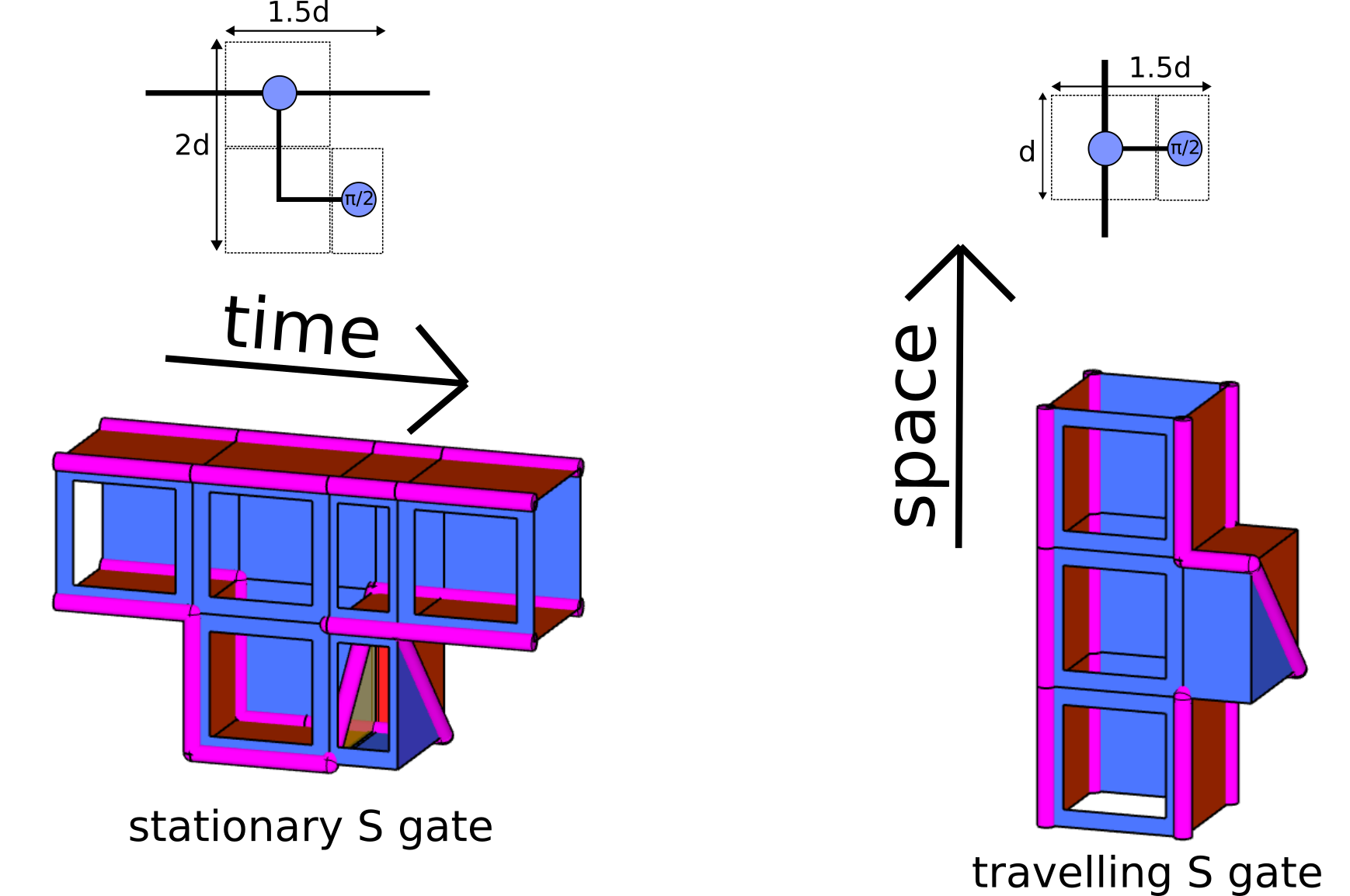}
    }
    \caption{
        ZX calculus diagrams and defect diagrams of using fast Y basis measurement to perform an S gate by gate teleportation~\cite{gottesman1999gateteleport}.
        In previous work, the S gate required $4 d^3$  volume~\cite{bombin2021logical,chamberland2022universal}.
        Here the S gate uses $3 d^3$ volume if applied to a stationary qubit, or $1.5d^3$ volume if applied while a qubit is being moved.
        The dependence of the cost on the motion of the logical qubit is due to the diagonal-twist-movement technique presented in this paper being specific to moving diagonally across space (as opposed to across time or spacetime).
        See \tbl{defect_types} for a summary of the types of defects appearing in this diagram.
        The windows cut into the front faces of the defects are just to show the interior; they aren't actually present in the construction.
    }
    \label{fig:other_improvements}
\end{figure}

It's notable that the timelike Y error mechanism shown in \fig{logical_error_paths} has distance $d$, despite being caused by errors spanning across $d/2$ rounds and not spanning across space.
This is because the relevant error mechanism requires a chain of X errors \emph{and} a chain of Z errors crossing the $d/2$ rounds.
Although there are error mechanisms that grow X and Z error chains together across space (e.g. Y basis data errors), there is apparently no corresponding mechanism that does this across time.
A consequence of this asymmetry between space and time is that, unlike most surface code constructions, the Y basis measurement has a preferred spacetime orientation.
Normally, performing a spacetime rotation of a defect diagram produces a different implementation of the same operation, with the same code distance.
The Y basis measurement is different: a spacetime rotation of its defect diagram will turn the timelike Y error mechanism into a spacelike Y error mechanism that has half the distance.
Avoiding this problem is why \fig{other_improvements} has two cases, one for spacelike motion and one for timelike motion.
It's conceivable that someone could fix this asymmetry by finding an implementation of the surface code cycle where the relevant spacelike Y error mechanisms had full distance, but I currently know of no way to do this.

\begin{figure}
    \centering
    \resizebox{\linewidth}{!}{
        \includegraphics{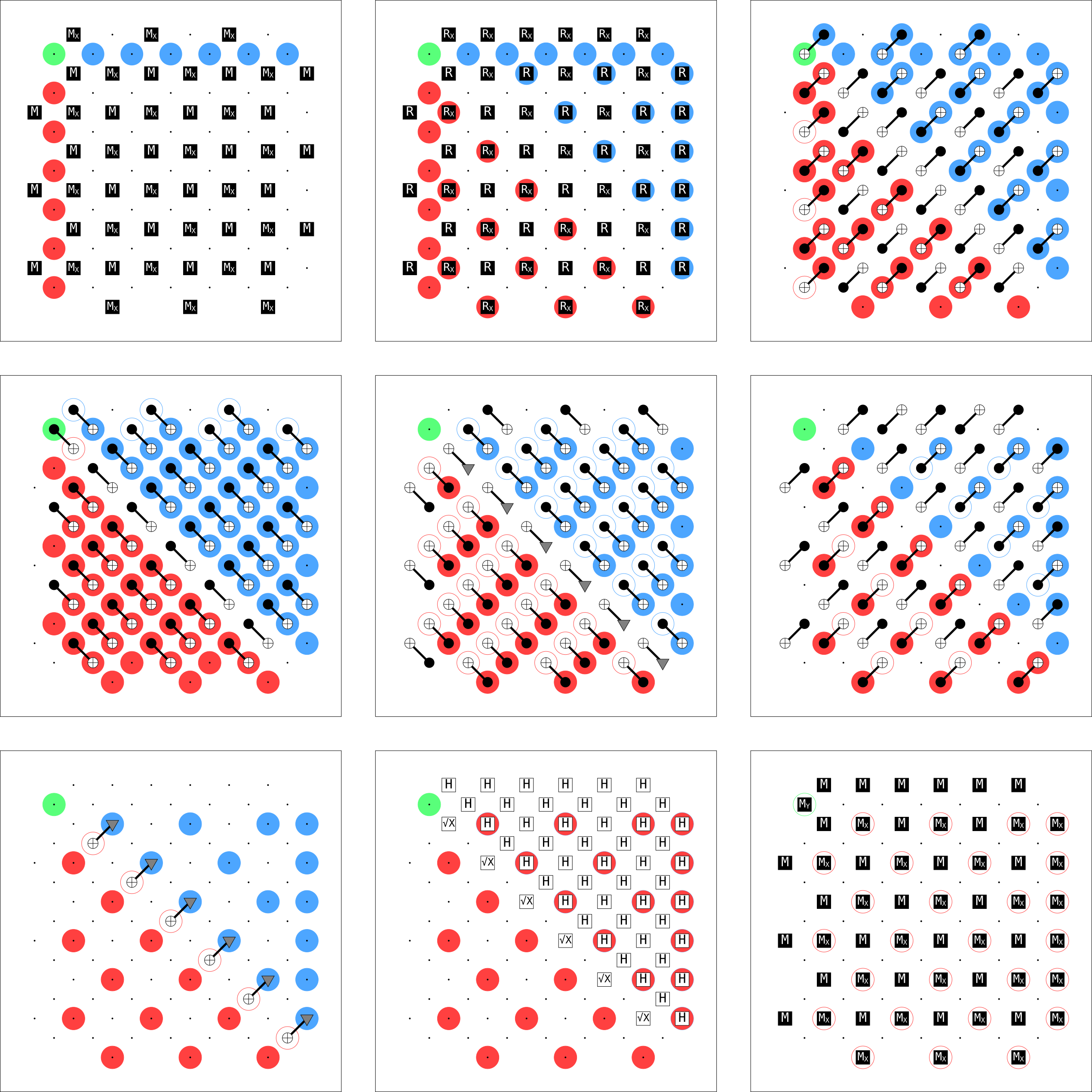}
    }
    \caption{
        Observable slice diagram of the surface code cycle where the logical qubit's Y observable is turned into a product of stabilizers.
        Filled circles are terms of the observable after the operations from that step have executed, using the color-to-pauli convention RGB=XYZ (red is X, green is Y, blue is Z).
        Unfilled circles are terms of the observable from before the operations from that step were executed.
        In the first step, the terms form a typical Y basis observable.
        In the second step, single qubit terms prepared by reset gates are added into the observable.
        The observable is then deformed by Clifford operations until, in the last step, all its terms disappear into measurements.
        Although the observable is now physically measured, more rounds of stabilizer measurement are needed in order to correct errors to make the logical measurement fault tolerant.
        \href{https://algassert.com/crumble\#circuit=Q(0,2)0;Q(0,4)1;Q(0,6)2;Q(0.5,0.5)3;Q(0.5,1.5)4;Q(0.5,2.5)5;Q(0.5,3.5)6;Q(0.5,4.5)7;Q(0.5,5.5)8;Q(0.5,6.5)9;Q(1,0)10;Q(1,1)11;Q(1,2)12;Q(1,3)13;Q(1,4)14;Q(1,5)15;Q(1,6)16;Q(1.5,0.5)17;Q(1.5,1.5)18;Q(1.5,2.5)19;Q(1.5,3.5)20;Q(1.5,4.5)21;Q(1.5,5.5)22;Q(1.5,6.5)23;Q(2,0)24;Q(2,1)25;Q(2,2)26;Q(2,3)27;Q(2,4)28;Q(2,5)29;Q(2,6)30;Q(2,7)31;Q(2.5,0.5)32;Q(2.5,1.5)33;Q(2.5,2.5)34;Q(2.5,3.5)35;Q(2.5,4.5)36;Q(2.5,5.5)37;Q(2.5,6.5)38;Q(3,0)39;Q(3,1)40;Q(3,2)41;Q(3,3)42;Q(3,4)43;Q(3,5)44;Q(3,6)45;Q(3.5,0.5)46;Q(3.5,1.5)47;Q(3.5,2.5)48;Q(3.5,3.5)49;Q(3.5,4.5)50;Q(3.5,5.5)51;Q(3.5,6.5)52;Q(4,0)53;Q(4,1)54;Q(4,2)55;Q(4,3)56;Q(4,4)57;Q(4,5)58;Q(4,6)59;Q(4,7)60;Q(4.5,0.5)61;Q(4.5,1.5)62;Q(4.5,2.5)63;Q(4.5,3.5)64;Q(4.5,4.5)65;Q(4.5,5.5)66;Q(4.5,6.5)67;Q(5,0)68;Q(5,1)69;Q(5,2)70;Q(5,3)71;Q(5,4)72;Q(5,5)73;Q(5,6)74;Q(5.5,0.5)75;Q(5.5,1.5)76;Q(5.5,2.5)77;Q(5.5,3.5)78;Q(5.5,4.5)79;Q(5.5,5.5)80;Q(5.5,6.5)81;Q(6,0)82;Q(6,1)83;Q(6,2)84;Q(6,3)85;Q(6,4)86;Q(6,5)87;Q(6,6)88;Q(6,7)89;Q(6.5,0.5)90;Q(6.5,1.5)91;Q(6.5,2.5)92;Q(6.5,3.5)93;Q(6.5,4.5)94;Q(6.5,5.5)95;Q(6.5,6.5)96;Q(7,1)97;Q(7,2)98;Q(7,3)99;Q(7,4)100;Q(7,5)101;Q(7,6)102;POLYGON(0,0,1,0.5)3_17_18_4;POLYGON(0,0,1,0.5)5_19_20_6;POLYGON(0,0,1,0.5)7_21_22_8;POLYGON(0,0,1,0.5)18_33_34_19;POLYGON(0,0,1,0.5)20_35_36_21;POLYGON(0,0,1,0.5)22_37_38_23;POLYGON(0,0,1,0.5)32_46_47_33;POLYGON(0,0,1,0.5)34_48_49_35;POLYGON(0,0,1,0.5)36_50_51_37;POLYGON(0,0,1,0.5)47_62_63_48;POLYGON(0,0,1,0.5)49_64_65_50;POLYGON(0,0,1,0.5)51_66_67_52;POLYGON(0,0,1,0.5)61_75_76_62;POLYGON(0,0,1,0.5)63_77_78_64;POLYGON(0,0,1,0.5)65_79_80_66;POLYGON(0,0,1,0.5)76_91_92_77;POLYGON(0,0,1,0.5)78_93_94_79;POLYGON(0,0,1,0.5)80_95_96_81;POLYGON(0,0,1,0.75)4_5;POLYGON(0,0,1,0.75)6_7;POLYGON(0,0,1,0.75)8_9;POLYGON(0,0,1,0.75)90_91;POLYGON(0,0,1,0.75)92_93;POLYGON(0,0,1,0.75)94_95;POLYGON(1,0,0,0.5)4_18_19_5;POLYGON(1,0,0,0.5)6_20_21_7;POLYGON(1,0,0,0.5)8_22_23_9;POLYGON(1,0,0,0.5)17_32_33_18;POLYGON(1,0,0,0.5)19_34_35_20;POLYGON(1,0,0,0.5)21_36_37_22;POLYGON(1,0,0,0.5)33_47_48_34;POLYGON(1,0,0,0.5)35_49_50_36;POLYGON(1,0,0,0.5)37_51_52_38;POLYGON(1,0,0,0.5)46_61_62_47;POLYGON(1,0,0,0.5)48_63_64_49;POLYGON(1,0,0,0.5)50_65_66_51;POLYGON(1,0,0,0.5)62_76_77_63;POLYGON(1,0,0,0.5)64_78_79_65;POLYGON(1,0,0,0.5)66_80_81_67;POLYGON(1,0,0,0.5)75_90_91_76;POLYGON(1,0,0,0.5)77_92_93_78;POLYGON(1,0,0,0.5)79_94_95_80;POLYGON(1,0,0,0.75)17_3;POLYGON(1,0,0,0.75)38_23;POLYGON(1,0,0,0.75)46_32;POLYGON(1,0,0,0.75)67_52;POLYGON(1,0,0,0.75)75_61;POLYGON(1,0,0,0.75)96_81;TICK;MARKX(0)4_5_6_7_8_9;MARKY(0)3;MARKZ(0)17_32_46_61_75_90;TICK;R_0_1_2_11_13_15_26_28_30_40_42_44_55_57_59_69_71_73_84_86_88_97_98_99_100_101_102;RX_10_12_14_16_24_25_27_29_31_39_41_43_45_53_54_56_58_60_68_70_72_74_82_83_85_87_89;MARKX(0)12_14_16_27_29_43_45_58_74_31_60_89;MARKZ(0)40_55_69_71_84_86_97_98_99_100_101_102;TICK;CX_10_3_12_5_14_7_16_9_25_18_27_20_29_22_39_32_41_34_43_36_45_38_54_47_56_49_58_51_68_61_70_63_72_65_74_67_83_76_85_78_87_80_4_11_6_13_8_15_19_26_21_28_23_30_33_40_35_42_37_44_48_55_50_57_52_59_62_69_64_71_66_73_77_84_79_86_81_88_91_97_93_99_95_101;TICK;CX_10_17_12_19_14_21_16_23_25_33_27_35_29_37_39_46_41_48_43_50_45_52_54_62_56_64_58_66_68_75_70_77_72_79_74_81_83_91_85_93_87_95_3_11_5_13_7_15_18_26_20_28_22_30_32_40_34_42_36_44_47_55_49_57_51_59_61_69_63_71_65_73_76_84_78_86_80_88_90_97_92_99_94_101;TICK;CX_17_25_33_41_46_54_48_56_62_70_64_72_75_83_77_85_79_87_91_98_93_100_95_102_24_32_40_47_53_61_55_63_69_76_71_78_82_90_84_92_86_94_12_4_14_6_16_8_27_19_29_21_31_23_43_35_45_37_58_50_60_52_74_66_89_81_5_0_7_1_9_2_20_13_22_15_36_28_38_30_51_44_67_59;XCY_11_18_26_34_42_49_57_65_73_80_88_96;TICK;CX_32_39_47_54_61_68_63_70_76_83_78_85_92_98_94_100_24_17_40_33_53_46_55_48_69_62_71_64_82_75_84_77_86_79_97_91_99_93_101_95_12_18_14_20_16_22_27_34_29_36_31_38_43_49_45_51_58_65_60_67_74_80_89_96_4_0_6_1_8_2_19_13_21_15_35_28_37_30_50_44_66_59;TICK;XCY_18_25_34_41_49_56_65_72_80_87_96_102;TICK;H_17_32_33_46_47_48_61_62_63_64_75_76_77_78_79_90_91_92_93_94_95_10_24_25_39_40_41_53_54_55_56_68_69_70_71_72_82_83_84_85_86_87_97_98_99_100_101_102;SQRT_X_11_26_42_57_73_88;TICK;M_0_1_2_10_11_13_15_24_26_28_30_39_40_42_44_53_55_57_59_68_69_71_73_82_84_86_88;MX_12_14_16_25_27_29_31_41_43_45_54_56_58_60_70_72_74_83_85_87_89_97_98_99_100_101_102;MY_3;MARKX(0)12_14_16_27_29_41_43_45_56_58_70_72_74_85_87_25_54_97_98_99_100_101_102_89_60_31_83;MARKY(0)3;TICK;POLYGON(0,0,1,0.5)17_18_4;POLYGON(0,0,1,0.5)5_19_20_6;POLYGON(0,0,1,0.5)7_21_22_8;POLYGON(0,0,1,0.5)18_33_34_19;POLYGON(0,0,1,0.5)20_35_36_21;POLYGON(0,0,1,0.5)22_37_38_23;POLYGON(0,0,1,0.5)32_46_47_33;POLYGON(0,0,1,0.5)34_48_49_35;POLYGON(0,0,1,0.5)36_50_51_37;POLYGON(0,0,1,0.5)47_62_63_48;POLYGON(0,0,1,0.5)49_64_65_50;POLYGON(0,0,1,0.5)51_66_67_52;POLYGON(0,0,1,0.5)61_75_76_62;POLYGON(0,0,1,0.5)63_77_78_64;POLYGON(0,0,1,0.5)65_79_80_66;POLYGON(0,0,1,0.5)76_91_92_77;POLYGON(0,0,1,0.5)78_93_94_79;POLYGON(0,0,1,0.5)80_95_96_81;POLYGON(0,0,1,0.75)4_5;POLYGON(0,0,1,0.75)6_7;POLYGON(0,0,1,0.75)8_9;POLYGON(0,0,1,0.75)32_17;POLYGON(0,0,1,0.75)61_46;POLYGON(0,0,1,0.75)90_75;POLYGON(1,0,0,0.5)4_18_19_5;POLYGON(1,0,0,0.5)6_20_21_7;POLYGON(1,0,0,0.5)8_22_23_9;POLYGON(1,0,0,0.5)17_32_33_18;POLYGON(1,0,0,0.5)19_34_35_20;POLYGON(1,0,0,0.5)21_36_37_22;POLYGON(1,0,0,0.5)33_47_48_34;POLYGON(1,0,0,0.5)35_49_50_36;POLYGON(1,0,0,0.5)37_51_52_38;POLYGON(1,0,0,0.5)46_61_62_47;POLYGON(1,0,0,0.5)48_63_64_49;POLYGON(1,0,0,0.5)50_65_66_51;POLYGON(1,0,0,0.5)62_76_77_63;POLYGON(1,0,0,0.5)64_78_79_65;POLYGON(1,0,0,0.5)66_80_81_67;POLYGON(1,0,0,0.5)75_90_91_76;POLYGON(1,0,0,0.5)77_92_93_78;POLYGON(1,0,0,0.5)79_94_95_80;POLYGON(1,0,0,0.75)38_23;POLYGON(1,0,0,0.75)67_52;POLYGON(1,0,0,0.75)96_81;POLYGON(1,0,0,0.75)91_92;POLYGON(1,0,0,0.75)93_94;POLYGON(1,0,0,0.75)95_96}{Click here to open this circuit in crumble.}
    }
    \label{fig:obs_slice}
\end{figure}

\section{Benchmarking}
\label{sec:benchmarking}

All circuits in this paper were simulated using the noise model defined in \app{noise_model}.
Decoding was done using an internal correlated minimum weight perfect matching decoder written by Austin Fowler.
For each circuit, samples were taken until either a thousand logical errors had been seen or a billion shots had been taken, whichever came first.
All simulated circuits were variations on memory experiments.
The code, circuits, and statistics are available \href{https://doi.org/10.5281/zenodo.7487893}{on Zenodo}~\cite{gidneyybasisdata2022}.

In a memory experiment, a known logical state is initialized, protected against noise for some number of rounds, and then measured.
The experiment succeeds if the measured state agrees with the prepared state, after error correction.
Access to a Y basis measurement (and, by time reversal, a Y basis initialization), allows a Y basis memory experiment to be defined.

\begin{figure}
    \centering
    \resizebox{\linewidth}{!}{
        \includegraphics{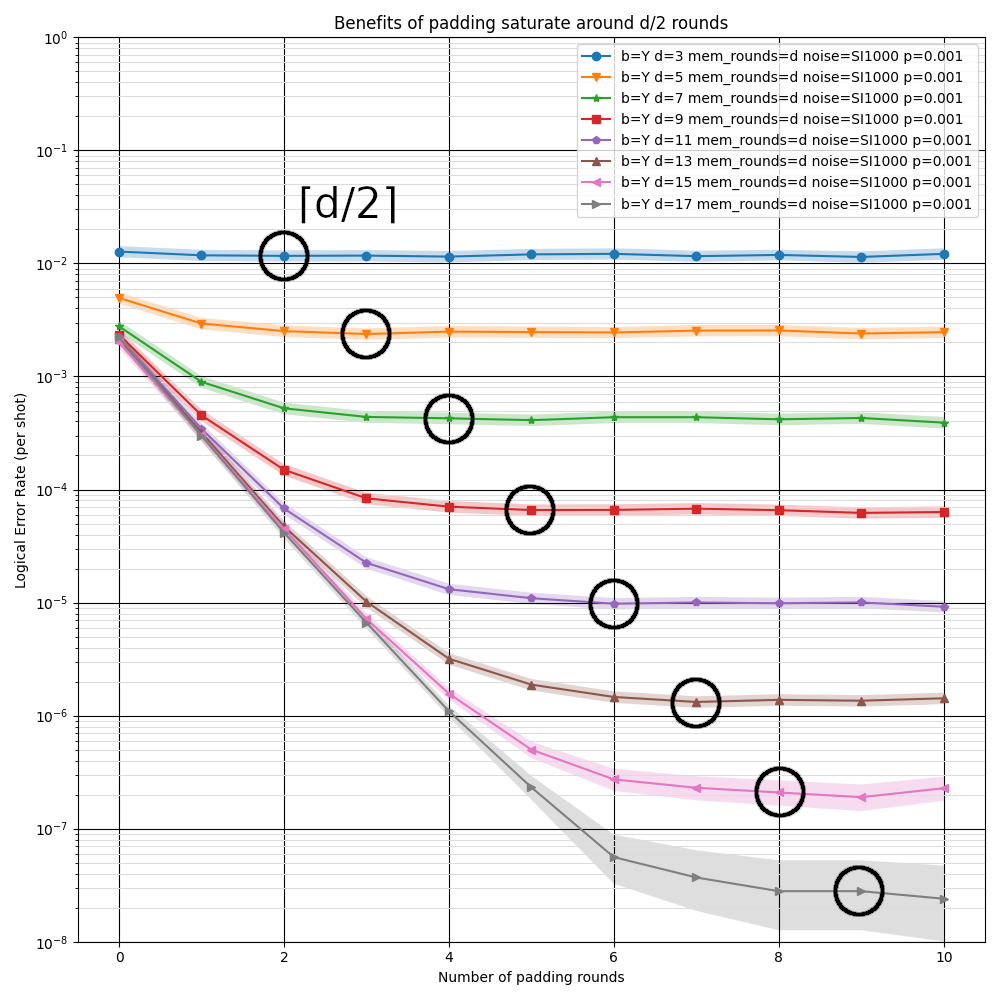}
    }
    \caption{
        Varying the number of padding rounds at various code distances in the Y basis memory experiment.
        When the number of padding rounds is too small, the error rate is limited by timelike errors.
        As the number of padding rounds is increased, the effect of timelike errors is exponentially suppressed until they become negligible relative to spacelike error mechanisms.
        The benefits of adding padding rounds saturates at around $d/2$ padding rounds.
        Color highlights show hypotheses with a likelihood within a factor of 1000 of the maximum likelihood hypothesis, given the samples collected.
    }
    \label{fig:pad_saturation}
\end{figure}

The Y basis memory experiment can be used to experimentally determine the optimal number of padding rounds to use for Y basis initialization and Y basis measurement.
If not enough padding rounds are used, timelike error mechanisms become dominant and limit the fidelity.
But the benefit of adding padding rounds saturates once these timelike error mechanisms have been suppressed below the noise floor set by spacelike error mechanisms.
\fig{pad_saturation} presents the results of simulations varying the number of padding rounds, showing that the benefit of adding padding rounds saturates at around $d/2$ rounds of padding.

An interesting property of the Y basis memory experiment is that preserving Y basis states (as opposed to X or Z basis states) better approximates the task of preserving the highly entangled states that would be present in a real quantum computation.
Because X and Z type defects form large 2D surfaces in spacetime, while Y type defects only form narrow 1D curves, there are more paths that chains of X type errors and Z type errors can follow.
Thus, logical errors in the surface code are biased away from being Y type errors, making X and Z type errors the dominant logical error mechanisms.
Y basis states and highly entangled states are sensitive to both of these dominant logical error mechanisms, while X and Z basis states are only sensitive to one.
This difference is especially clear when focusing on stabilizer dependence.
Although an X basis memory experiment measures both X type and Z type stabilizers, the logical value can be error corrected using only the measurement results from the X type stabilizers.
Similarly, a Z basis memory experiment can be error corrected using only the measurement results from the Z type stabilizers.
Protecting a Y basis state, or a highly entangled state, requires both the X stabilizer measurements and the Z stabilizer measurements.
The Y basis memory experiment exercises the whole surface code, instead of only half.

\begin{figure}
    \centering
    \resizebox{\linewidth}{!}{
        \includegraphics{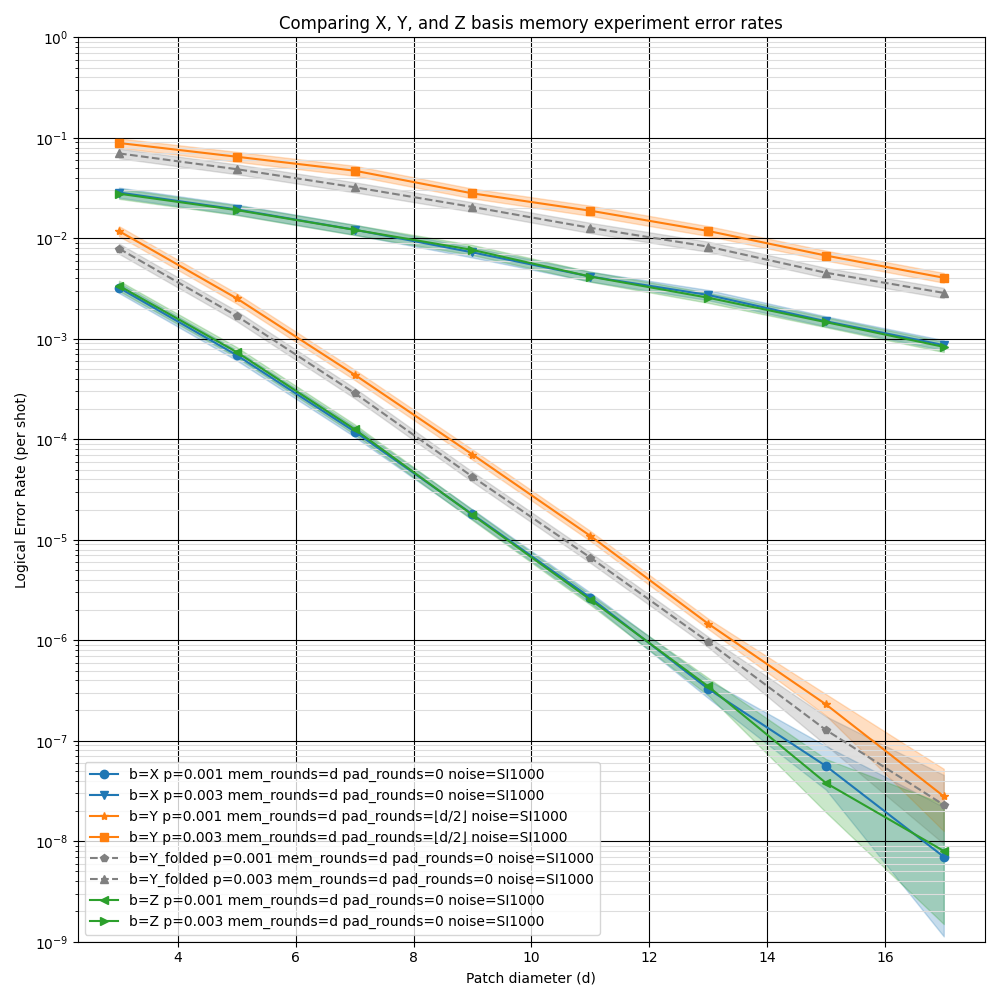}
    }
    \caption{
        Comparing the Y basis memory experiment to X and Z basis memory experiments, at various code distances and for two different physical noise strengths.
        When using the initialization and measurement presented in this paper, the Y basis memory experiment improves with code distance roughly as quickly as the X and Z basis memory experiments but has a starting logical error rate that is roughly five times higher.
        Using the folded surface code to initialize and measure in the Y basis, using non-local connectivity, would improve the logical error rate.
        Color highlights show hypotheses with a likelihood within a factor of 1000 of the maximum likelihood hypothesis, given the samples collected.
    }
    \label{fig:error_rate}
\end{figure}

Because the Y basis memory experiment is sensitive to both X and Z logical errors, I expected Y basis memory experiments to have logical error rates roughly twice as high as X and Z memory experiments.
However, simulations showed that the logical error rate was actually an additional 1.5 to 2.5 times higher than this expectation (see \fig{error_rate}).
I was surprised by this difference, so I investigated where the additional error was coming from.

\begin{figure}
    \centering
    \resizebox{\linewidth}{!}{
        \includegraphics{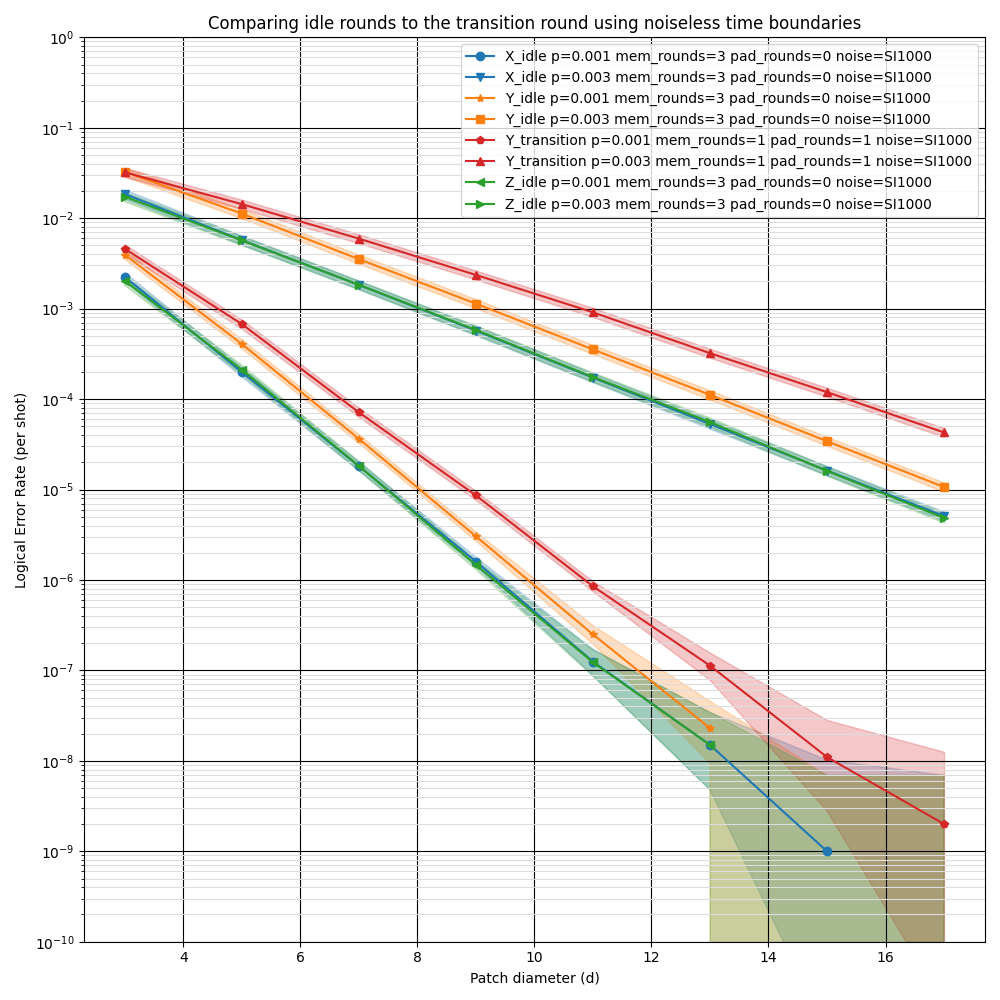}
    }
    \caption{
        Logical error rate during three rounds including a Y basis transition, versus three rounds of idling in various bases, with magically noiseless time boundaries before and after the rounds.
        Color highlights show hypotheses with a likelihood within a factor of 1000 of the maximum likelihood hypothesis, given the samples collected.
    }
    \label{fig:round_error_rate}
\end{figure}

Because of \fig{pad_saturation}, I knew timelike errors weren't the source of the problem.
If they were, adding more padding would have fixed the issue.
This left spacelike errors during the transition round as the likely cause.
I checked this by isolating the transition round using magically noiseless time boundaries.
(A magically noiseless time boundary performs perfect measurement of all stabilizers and observables of the code, preventing all error chains from crossing.)
As controls, I also used magically noiseless time boundaries to isolate three rounds of idling in the X, Y, and Z bases.
The results are shown in \fig{round_error_rate}.
They suggest that the transition round is the source of the additional errors.
(Also, they confirm Y basis idling is twice as hard as X or Z basis idling.)

\begin{figure}
    \centering
    \resizebox{0.65\linewidth}{!}{
        \includegraphics{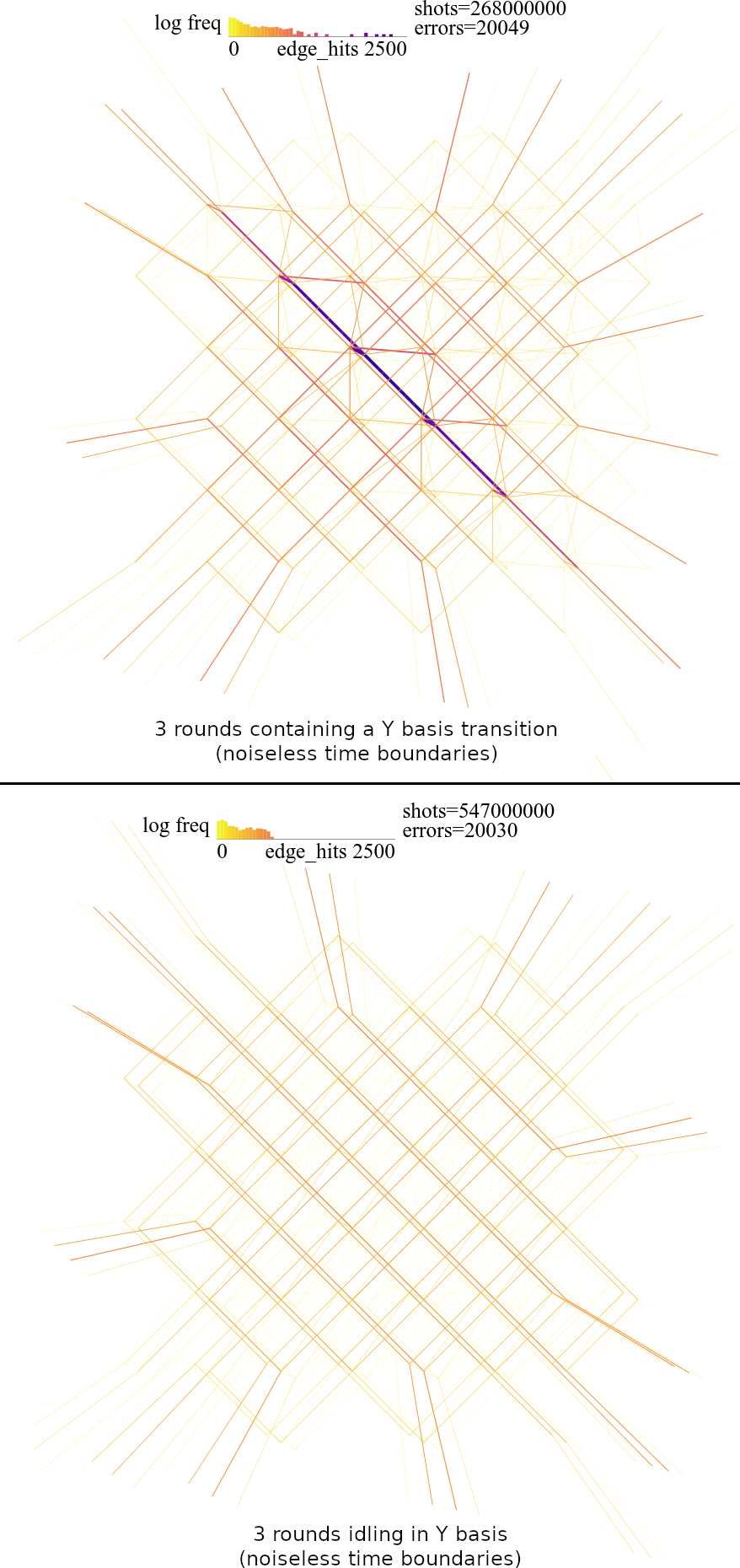}
    }
    \caption{
        Heatmaps of decoding graph edges contributing to logical errors from the $d=7$, $p=0.1\%$ case of \fig{round_error_rate}.
        Edges are highlighted relative to how often they appeared in logical errors (as a difference between the simulated noise's edges and the edges predicted by the decoder, keeping only the edges in connected components that flip the logical observable).
        Boundary edges have had their hit count artificially halved, to compensate for symptom degeneracy at the sides of the patch.
        Logical errors appear to be traveling along the twist during the Y basis transition.
    }
    \label{fig:error_heat}
\end{figure}

\fig{error_heat} digs further into the source of the extra errors by highlighting the paths that logical errors tend to follow during the transition round.
It suggests that the problem is occurring along the path of the crossing twist.
This makes sense to me, because the detectors along that path have the largest detecting regions (note where the biggest shapes are in \fig{transition_detector_slices}).
This suggests that reducing the size of those detecting regions could bring the logical error rate in line with expectation.
This is a place where there's room for improvement in the circuit construction.

\begin{figure}
    \centering
    \resizebox{0.8 \linewidth}{!}{
        \includegraphics{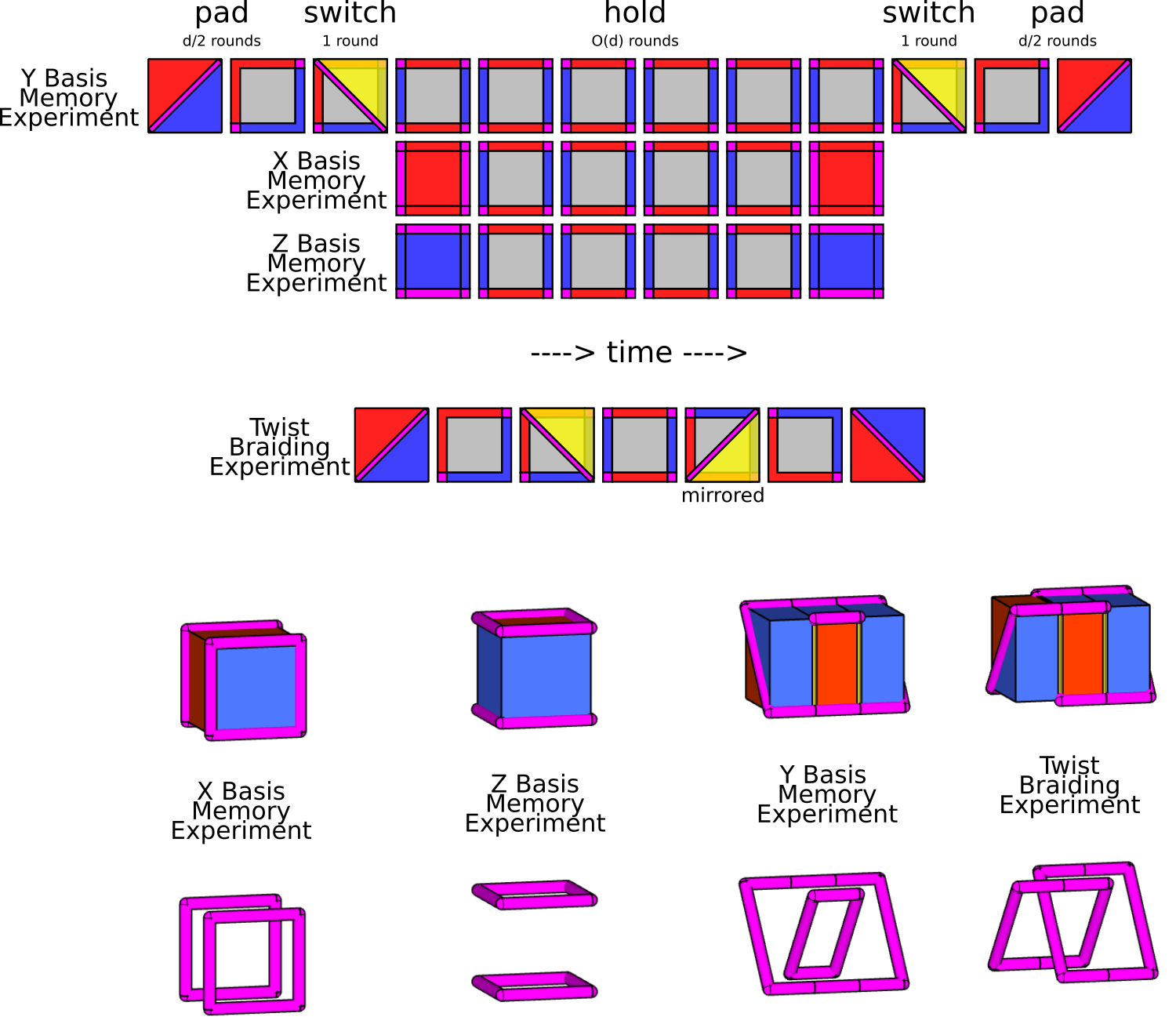}
    }
    \caption{
        Defect diagrams summarizing experiments simulated in this paper.
        See \tbl{defect_types} for a summary of the types of defects appearing in this diagram.
    }
    \label{fig:experiment_defect_diagrams}
\end{figure}

The last experiment that I simulated is an interesting variant of the Y basis memory experiment.
In the normal Y basis memory experiment, which I'll now start calling ``the unlinked experiment" to avoid ambiguity, the initialization and measurement use the same orientation for the crossing twist.
This produces a topology where the Y type defects form two unlinked rings in spacetime.
In the other variant, which I'll call ``the braiding experiment", the initialization and measurement use opposite orientations, resulting in a topology where the Y type defects form linked rings (see \fig{experiment_defect_diagrams}).

\begin{figure}
    \centering
    \resizebox{\linewidth}{!}{
        \includegraphics{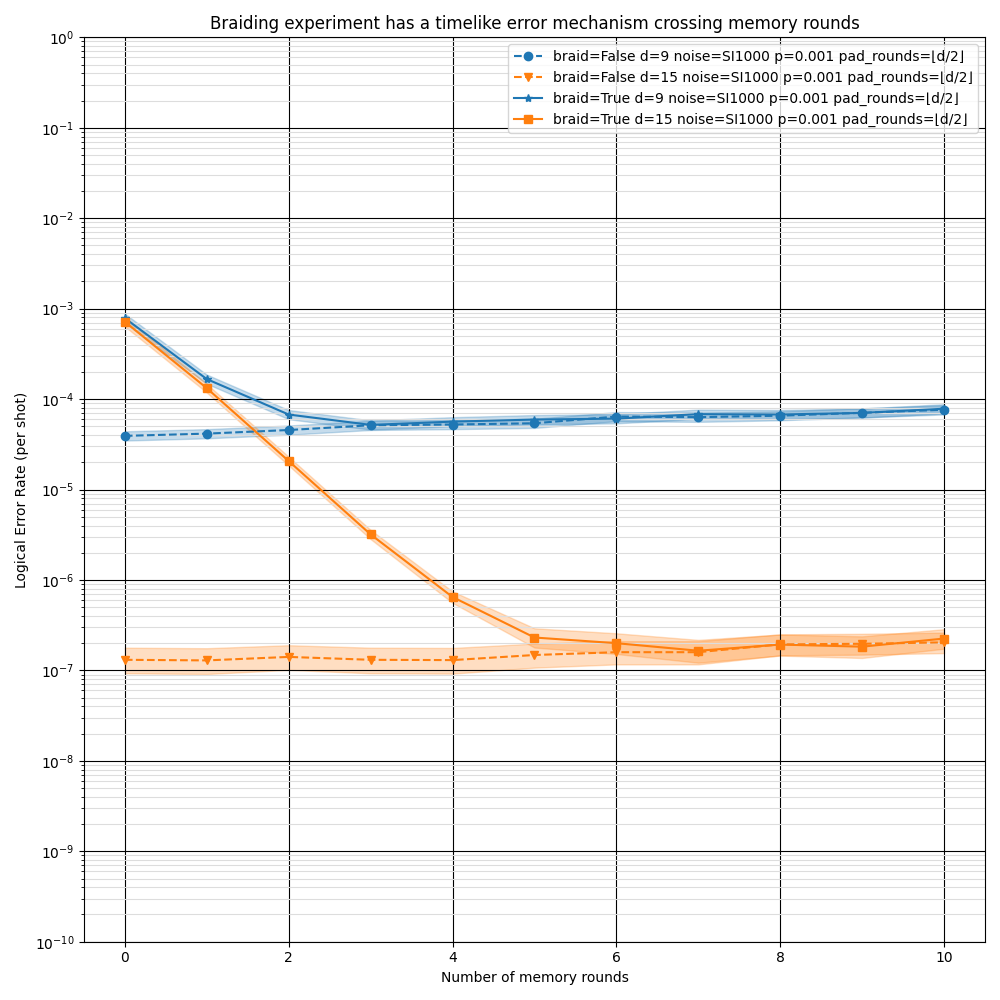}
    }
    \caption{
        Varying the number of rounds spent idling in the memory configuration in the two variations of the Y basis memory experiment.
        At small numbers of idling rounds, a timelike error in the braiding experiment between the initialization and the measurement is dominant, causing an exponential jump in logical errors as idling rounds approaches zero.
        Color highlights show hypotheses with a likelihood within a factor of 1000 of the maximum likelihood hypothesis, given the samples collected.
    }
    \label{fig:braiding_too_fast_error_rate}
\end{figure}

\fig{braiding_too_fast_error_rate} shows a key difference between the unlinked experiment and the braiding experiment.
In the unlinked experiment, it's always beneficial to reduce the time spent between the initialization and the measurement, whereas in the braiding experiment spending too few rounds idling increases the error rate dramatically.
This is because, in the unlinked experiment, the twist crossing the patch during initialization is part of the same spacetime object as the twist crossing the patch during measurement.
An error chain traveling from an object to itself has no effect on the state of the system (it is, effectively, a product of stabilizers).
In the braiding experiment, the twist crossing the patch during initialization is part of a different spacetime object than the twist crossing the patch during measurement.
It's a logical error for error chains to travel between these two objects, and if there are too few idling rounds separating the objects then this error mechanism begins to dominate.

The braiding experiment highlights a subtle caveat on my claim that it takes $d/2 + O(1)$ rounds to perform a Y basis measurement.
There are situations where up to $d/2 + O(1)$ additional idling rounds are needed, before the measurement, in order to keep the crossing twist away from other twists.
This context-dependent-cost phenomenon isn't unique to Y basis measurements.
For example, the spacetime volume of a CNOT operation is reported based on a specific starting orientation of the involved qubits~\cite{horsman2012latticesurgery}.
Other orientations require more volume.
The sheer variety of possible contexts, and in particular the complex interdependence between how qubits are routed and oriented and operated upon, mean that describing the cost of a single operation using a single number is unavoidably an approximation.
Ultimately, there are situations where the marginal cost of a Y basis measurement is in fact $d/2 + O(1)$ rounds and, based on my anecdotal experience assembling large surface code computations, I expect that these situations will be by far the most common.
Therefore, despite the inherent ambiguity, I think $d/2 + O(1)$ rounds is the most appropriate number to use to describe the time it takes to perform the Y basis measurement presented in this paper.

\begin{figure}
    \centering
    \resizebox{0.5 \linewidth}{!}{
        \includegraphics{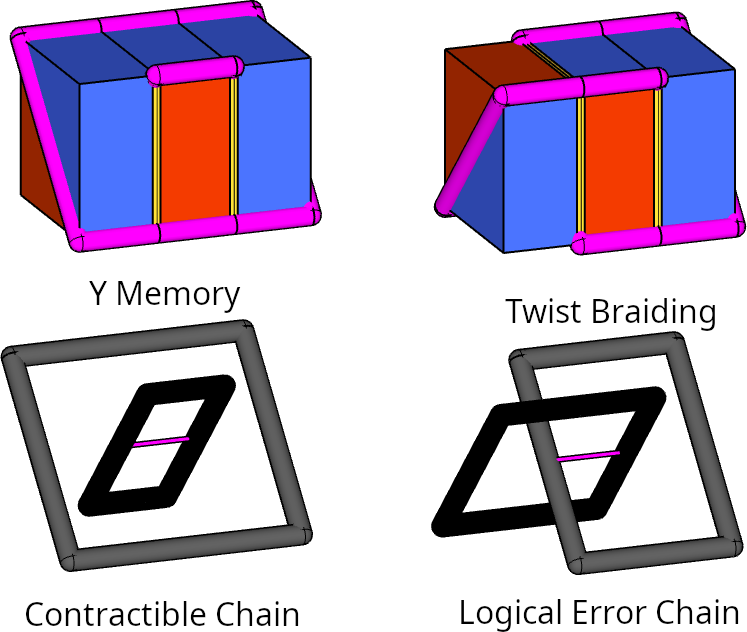}
    }
    \caption{
        Flipping the orientation of the Y measurement, relative to the Y basis initialization, produces a topologically distinct surface code experiment.
        In spacetime, the two twist defects end up either unlinked (the ``Y basis memory experiment'') or linked (the ``twist braiding experiment'').
        A series of X and Z measurement errors at the center of the patch, spanning from the preparation transition round to the measurement transition round, creates a logical error in the twist braiding experiment but is topologically vacuous in the Y memory experiment.
    }
    \label{fig:contractible_vs_topological_error}
\end{figure}

\section{Conclusion}
\label{sec:conclusion}

In this paper, I presented a way to perform Y basis measurement and Y basis initialization in $d/2 + O(1)$ rounds, without leaving the bounding box of the surface code patch and while achieving full code distance.
The spacetime volume of this construction is nearly an order of magnitude lower than achieved in previous work.
I benchmarked the performance of this construction, and found that in short memory experiments it produces logical error rates slightly higher than would be expected given the performance of X and Z basis memory experiments, but good enough for practical use.

Having a faster Y basis measurement allows other computational primitives to be improved.
For example, \fig{other_improvements} has concept diagrams for performing S gates by gate teleportation using the new Y basis measurement.
These S gates use less spacetime volume than previous work.
Faster Y basis measurement is also useful in the construction of magic state factories.
For example, the magic state factories in \cite{fowler2018latticesurgery,gidney2019catalyzeddistillation} end with a layer of potential Y basis measurements.
These measurements can now be performed faster than previously thought, increasing the expected execution speeds of those factories by around 10\%.

The fact that the Y basis measurement can be performed inplace unlocks the ability to do tomography on hardware only large enough to store a single logical qubit.
For example, suppose you're interested in experimentally validating the fidelity of $|T\rangle$ state injection.
If only X and Z basis measurement are available, it's impossible to distinguish between $|T\rangle$ and $|T^\dagger\rangle$, or between $|T\rangle$ and a partially dephased $|+\rangle$ state, because all these states have the same projection onto the XZ plane of the Bloch sphere.
Adding the ability to do Y basis measurement makes these states distinguishable.

Looking beyond the Y basis, I'm hopeful there are other uses for the ability to move a twist diagonally across a surface code patch.
More generally, I look forward to the discovery of more ways to lay defects across space and time, in ever more complex computational dances.

\section{Acknowledgements}

I thank Ben Brown and Cody Jones for discussions about S gates and $|i\rangle$ states that inspired the ideas that eventually turned into the Y basis measurement presented in this paper.
I thank Matt McEwen for extensive discussions on the construction, and for encouragement to continue improving upon earlier versions of the circuit.
I thank Austin Fowler for writing the correlated minimum weight perfect matching decoder used by this paper.
I thank Austin Fowler, Alexis Morvan, Mike Newman, and Kevin Satzinger for giving feedback that improved the paper.
I thank Hartmut Neven for creating an environment where this research was possible.

\printbibliography

\appendix

\section{Surface Code Defects}
\label{app:defects}

\begin{table}
    \centering
    \resizebox{\linewidth}{!}{
    \begin{tabular}{|c|c|c|c|c|c|c|c}
         \hline
         Defect Type & Name & Shape & Effect & In Diagrams \\
         \hline
         I & Bulk & 3D & Conserves X and Z detection event parity & Empty Space \\
         X & X Boundary & 2D & Absorbs $X$ detection events & Red Walls \\
         Z & Z Boundary & 2D & Absorbs $Z$ detection events & Blue Walls \\
         H & Domain Wall & 2D & Cross-links $X \xleftrightarrow{} Z$ detection events across the wall & Transparent Yellow Walls\\
         Y & Twist & 1D & Absorbs adjacent $X \cdot Z$ detection events & Magenta Cylinders \\
         \hline
    \end{tabular}
    }
    \caption{
        Summary of types of defects in surface code circuits.
    }
    \label{tbl:defect_types}
\end{table}

In quantum error correcting codes, a ``detector'' is a set of measurements that will produce the same parity every time when a circuit is executed without noise.
A ``detection event'' occurs when the parity of a detector is wrong.
Detection events in the surface code can be classified into two types (X and Z), based on whether they are associated with an X type or Z type stabilizer changing unexpectedly.
In the bulk of a surface code circuit, all errors produce an even number of detection events of each type.
In other words, the bulk of the surface code ``conserves X parity" and ``conserves Z parity".
These two conservation properties are what \emph{define} the bulk.
Any circuit location where Pauli errors conserve X detection event parity and conserve Z detection event parity is part of the bulk, regardless of the specific implementation details of the circuit.

Defects are circuit locations where X parity conservation and/or Z parity conservation is broken, because errors can produce an odd number of X detection events and/or an odd number of Z detection events.
Important types of defects are summarized in \tbl{defect_types}.
Locations in the circuit where an odd number of X detection events can be produced by an error are ``X type defects" or ``X boundaries".
Locations where an odd number of Z detection events can be produced by an error are ``Z type defects" or ``Z boundaries".
Locations where an error can produce one X detection event paired with one Z detection event, breaking the individual X and Z parities but not necessarily the combined $X+Z$ parity, are ``H type defects" or ``Y type defects".

The distinction between H type defects (``domain walls") and Y type defects (``twists") is whether or not it's possible to use the defect to attach a chain of X errors to a chain of Z errors arriving from the same direction.
An H type defect can't be used for this, because the relevant errors always place the relevant X/Z detection events on opposite sides of the wall, but a Y type defect can.
Note that, when a domain wall ends in the bulk, it always ends on a Y type defect.
An X chain passing near the end of a domain wall can attach to a Z chain at the same location by passing through the domain wall to become Z type and then returning by passing around the end of the wall.

There are many different ways to implement defects.
For example, in the unrotated surface code, a spatial X boundary is implemented by three body stabilizers whereas, in the rotated surface code, two body stabilizers are used.
Another example: the bulk of the surface code can be implemented on a square grid or on a hex grid~\cite{mcewenmidoutsurfaces2023}.
At a low level, the specific implementation used is extremely important because it affects performance (e.g. it can change the threshold).
At a high level, the different implementations of defects are interchangeable.
The computation performed by a surface code circuit can be understood in terms of the layout and topology of its spacetime defects, without knowing exactly how the defects are implemented.
Thus, surface code computations can be improved both macroscopically, by changing the arrangement of defects, and microscopically, by changing the implementation of defects.

\section{Noise Model}
\label{app:noise_model}

\begin{figure}
    \centering
    \resizebox{\linewidth}{!}{
        \includegraphics{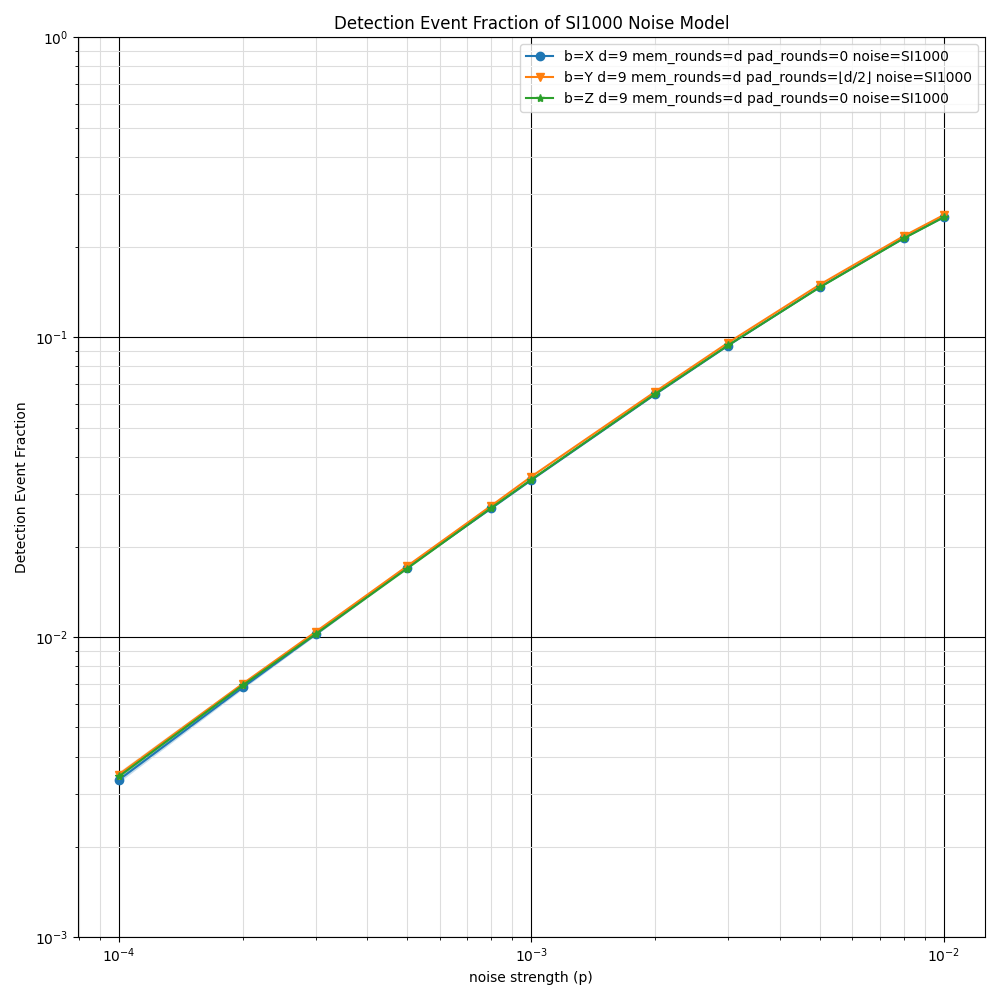}
    }
    \caption{
        Detection event fraction of basic memory experiment circuits in the SI1000 noise model.
        The detection event fraction is the probability of a detector producing a detection event, averaged over all detectors in the circuit.
    }
    \label{fig:det_frac}
\end{figure}

All circuits in this paper were simulated using the superconducting-inspired noise model defined in \tbl{noise_model}.
In legends, this noise model is called ``SI1000" (short for Superconducting Inspired with 1000 nanosecond cycle).
To provide a reference for comparisons to other error models or to hardware, the detection event fraction of the model at various noise strengths is plotted in \fig{det_frac}.

\begin{table}
    \centering
    \begin{tabular}{|r|l|}
    \hline
    Noise channel & Probability distribution of effects
    \\
    \hline
    $\text{MERR}(p)$ & $\begin{aligned}
        1-p &\rightarrow \text{(report previous measurement correctly)}
        \\
        p &\rightarrow \text{(report previous measurement incorrectly; flip its result)}
    \end{aligned}$
    \\
    \hline
    $\text{XERR}(p)$ & $\begin{aligned}
        1-p &\rightarrow I
        \\
        p &\rightarrow X
    \end{aligned}$
    \\
    \hline
    $\text{ZERR}(p)$ & $\begin{aligned}
        1-p &\rightarrow I
        \\
        p &\rightarrow Z
    \end{aligned}$
    \\
    \hline
    $\text{DEP1}(p)$ & $\begin{aligned}
        1-p &\rightarrow I
        \\
        p/3 &\rightarrow X
        \\
        p/3 &\rightarrow Y
        \\
        p/3 &\rightarrow Z
    \end{aligned}$
    \\
    \hline
    $\text{DEP2}(p)$ & $\begin{aligned}
        1-p &\rightarrow I \otimes I
        &\;\;
        p/15 &\rightarrow I \otimes X
        &\;\;
        p/15 &\rightarrow I \otimes Y
        &\;\;
        p/15 &\rightarrow I \otimes Z
        \\
        p/15 &\rightarrow X \otimes I
        &\;\;
        p/15 &\rightarrow X \otimes X
        &\;\;
        p/15 &\rightarrow X \otimes Y
        &\;\;
        p/15 &\rightarrow X \otimes Z
        \\
        p/15 &\rightarrow Y \otimes I
        &\;\;
        p/15 &\rightarrow Y \otimes X
        &\;\;
        p/15 &\rightarrow Y \otimes Y
        &\;\;
        p/15 &\rightarrow Y \otimes Z
        \\
        p/15 &\rightarrow Z \otimes I
        &\;\;
        p/15 &\rightarrow Z \otimes X
        &\;\;
        p/15 &\rightarrow Z \otimes Y
        &\;\;
        p/15 &\rightarrow Z \otimes Z
    \end{aligned}$
    \\
    \hline
    \end{tabular}
    \caption{
        Definitions of various noise channels.
        Used by \tbl{noise_model}.
    }
    \label{tbl:noise_channels}
\end{table}

\begin{table}
    \centering
    \begin{tabular}{|r|l|}
    \hline
    Ideal gate & Noisy gate
    \\
    \hline
    (any single qubit unitary, including idling) $U_1$ & $U_1 \circ \text{DEP1}(p / 10)$
    \\
    $CZ$ & $CZ \circ \text{DEP2}(p)$
    \\
    \hline
    $R_Z$ & $R_Z \circ \text{XERR}(2p)$
    \\
    $M_Z$ & $M_Z \circ \text{MERR}(5p) \circ \text{DEP1}(p)$
    \\
    \hline
    (Wait for $M_Z$ or $R_Z$) & $\text{DEP1}(2p)$
    \\
    \hline
    \end{tabular}
    \caption{
        The superconducting-inspired noise model used by simulations in this paper.
        Same as ``SI1000" from \cite{gidney2021honeycombmemory}.
        The single parameter $p$ sets the two qubit gate error rate, with other error rates being relative to this rate.
        Measurements are noisiest while single qubit gates are least noisy.
        Qubits not being reset or measured during layers containing measurements or resets incur additional depolarization on top of other error mechanisms.
        Note $A \circ B = B \cdot A$ means $B$ is applied after $A$.
        Noise channels are defined in \tbl{noise_channels}.
    }
    \label{tbl:noise_model}
\end{table}

\section{Tactics for building complex QEC circuits}
\label{app:the_struggle}

In the past, when I've built quantum error correction circuits, I've used a strategy based on writing code to produce the desired circuit and then verifying that the final circuit was correct (e.g. by using Stim~\cite{gidney2021stim} to verify the circuit's graphlike code distance is as expected).
For the Y basis circuits presented in this paper, that approach didn't work.
There were two major obstacles.
First, although I could see roughly what had to happen in the circuit, I found it very difficult to work out exactly what the circuit should look like along the twist and at the boundaries of the patch.
Second, even once I knew what the circuit should look like, it was difficult to write correct code for producing the circuit due to there being so many opportunities for easy-to-make mistakes like off-by-one errors and sign errors.
Because of the these two obstacles, I would often get stuck in loops of fixing a problem by introducing a different problem.

\begin{figure}
    \centering
    \resizebox{\linewidth}{!}{
    \includegraphics{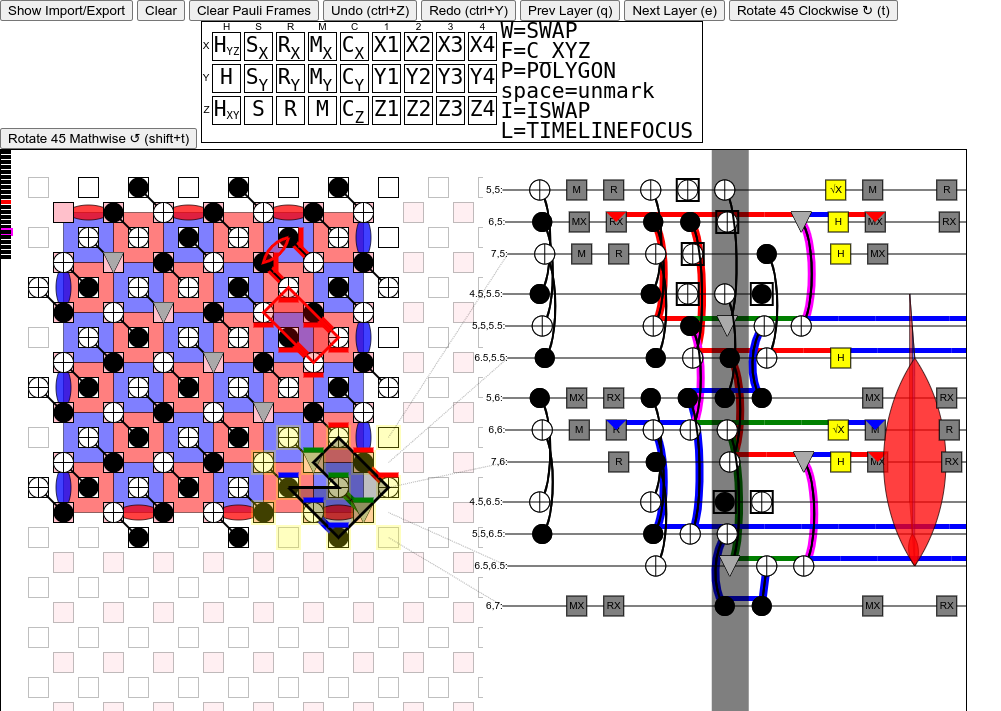}
    }
    \caption{
        Screenshot of ``Crumble", a stabilizer circuit editor that allows Pauli products to be interactively introduced and automatically propagated.
        Although Crumble is still in the prototyping stage, with core functionality unfinished and a complete lack of user interface polish, it was crucial to my process of creating and understanding the circuits described in this paper.
        A prototype version of Crumble can be accessed at \href{https://algassert.com/crumble}{algassert.com/crumble}.
        The circuits described in this paper can be loaded into Crumble by downloading the circuit files from Zenodo~\cite{gidneyybasisdata2022} and using Crumble's Import/Export button to paste and load the contents of a circuit into the editor.
    }
    \label{fig:crumble}
\end{figure}

To make it easier to figure out what the circuit should look like, I wrote a stabilizer circuit editor with interactive automatic Pauli product propagation (see \fig{crumble}).
This was helpful for two reasons.
First, tracking Pauli observables is a very common task when designing QEC circuits, but it's slow and error-prone to do by hand.
Computer software can do it billions of times faster, and more reliably.
Second, manually editing the circuit is a much more direct form of interaction than iterating on code for producing a circuit~\cite{brettvictormediaforthinking2013}.
This completely nullifies whole classes of errors.
For example, I make off-by-one errors when placing gates by math but I don't make these errors when placing gates by hand.
The key circuit cycle in \fig{transition_detector_slices} was built, end to end, in this interactive editor before being translated into code.

To overcome the fact that each piece of the circuit was hard to get right, I switched from an integration testing paradigm to a unit testing paradigm.
I focused on verifying that each individual surface code cycle was correct before attempting to combine it with the other cycles.
Each cycle was still defined by a circuit, but now each cycle also had an accompanying list of required ``stabilizer flows".
A stabilizer flow $A \xrightarrow{m} B$ has a desired starting stabilizer $A$, an expected ending stabilizer $B$, and a set of measurements $m$ to multiply into the stabilizer as it travels across the circuit.
For example, if a cycle was supposed to measure a stabilizer $S$ using measurements $m_1$ and $m_2$, then its list of stabilizer flows would contain $S \xrightarrow{m_1, m_2} 1$.
This allows each piece of the circuit to be independently verified because, given a circuit and a list of stabilizer flows, it's possible to mechanically check whether or not the circuit has all the flows.
(Note from the future: I ported my method for verifying flows into Stim. It can be used by calling \href{https://github.com/quantumlib/Stim/blob/main/doc/python_api_reference_vDev.md#stim.Circuit.has_all_flows}{\path{stim.Circuit.has_all_flows}}.)

As a result of using this approach to piece verification, I found myself undergoing an interesting conceptual shift.
I stopped thinking of surface code rounds as \emph{preserving} stabilizers, and started thinking of them as \emph{recreating} stabilizers.
Instead of specifying $A \rightarrow A$ and $A \xrightarrow{m} 1$ (translating to ``$A$ is preserved and $A$ is measured"), I found it more useful to specify $A \xrightarrow{m} 1$ and $1 \xrightarrow{m} A$ (translating to ``$A$ is measured and $A$ is prepared").
The benefit of this approach is that, when composing circuits, their flows can be matched together in order to form detectors.
This turns annotating detectors, which I previously found to be one of the most error prone integration processes, into an automatic process.

\end{document}